\pdfoutput=1
\documentclass[PAPER, atlasdraft=false, cernpreprint, UKenglish, texmf, orcidlogo]{atlasdoc}
\usepackage{atlaspackage}
\usepackage{atlasbiblatex}
\usepackage{multirow}
\usepackage{array}
\usepackage{xcolor}
\usepackage{bbding}
\usepackage{pifont}
\usepackage{wasysym}
\usepackage{lscape}
\usepackage{booktabs}
\usepackage{dcolumn}
\usepackage{comment}
\usepackage{rotating}
\newcolumntype{~}{>{\global\let\currentrowstyle\relax}}
\newcolumntype{^}{>{\currentrowstyle}}
\newcommand{\rowstyle}[1]{\gdef\currentrowstyle{#1}%
#1\ignorespaces
}

\usepackage{atlasphysics}
\graphicspath{{logos/}{figures/}}

\usepackage{ANA-EXOT-2021-34-PAPER-defs}

\AtlasTitle{Search for Beyond the Standard Model physics with anomaly detection in multilepton final states in $pp$ collisions at $\sqrt{s}=13~\tev$ with the ATLAS detector}

\AtlasAbstract{%
A model-agnostic search for Beyond the Standard Model physics is presented, targeting final states with at least four light leptons (electrons or muons). The search regions are separated by event topology and unsupervised machine learning is used to identify anomalous events in the full 140 fb$^{-1}$ of proton--proton collision data collected with the ATLAS detector during Run 2.
No significant excess above the Standard Model background expectation is observed.
Model-agnostic limits are presented in each topology, along with limits on several benchmark models including vector-like leptons, wino-like charginos and neutralinos, or smuons. Limits are set on the flavourful vector-like lepton model for the first time.
}

\AtlasRefCode{EXOT-2021-34}

\PreprintIdNumber{CERN-EP-2025-189}

\AtlasJournal{EPJC}
\AtlasJournalRef{\EPJC 86 (2026) 247}
\AtlasDOI{https://doi.org/10.1140/epjc/s10052-025-15218-2}

\hypersetup{pdftitle={ATLAS document},pdfauthor={The ATLAS Collaboration}}

\begin{document}

\maketitle

\tableofcontents

\section{Introduction}
\label{sec:intro}

Throughout the data taking years of 2015--2018, the ATLAS detector at the LHC has collected proton--proton collisions amassing \lumival of data.
The Standard Model (SM) of particle physics has been able to describe collider physics phenomena to outstanding precision. Its last missing component, the Higgs boson, was discovered in 2012 by the ATLAS and CMS Collaborations~\cite{HIGG-2012-27,CMS-HIG-12-028}.
However, the SM remains an incomplete theory, as it does not provide answers to open questions such as the hierarchy of masses of elementary fermions, the origin of neutrino masses, the hierarchy and fine-tuning problems, the observed baryon asymmetry in the universe, and the nature of dark matter and dark energy. Many beyond-the-SM (BSM) theories have been proposed to address these and other shortcomings of the SM; however there are no clear indications of which BSM theory (if any) would provide a more accurate description of nature. Consequently the ATLAS experiment continues a broad BSM search program.
The traditional route for searching for BSM physics is to design an analysis targeting a specific model that results in detectable signatures at the LHC\@. However, without a preferred BSM physics model or scale, model-dependent searches will miss signatures of some possible BSM scenarios.
This analysis looks to solve this problem in collision events with multiple leptons, by casting a wide net to simultaneously cover a range of possible final states.

High lepton multiplicities are particularly interesting as they provide a comparatively low-SM-background channel to search for potential BSM physics.
ATLAS has previously performed searches with minimal model dependence in the multilepton final states~\cite{EXOT-2012-20, EXOT-2019-36}.
The search presented here is focused on light leptons (electrons, $e$, or muons, $\mu$). Tau leptons are allowed inclusively but not explicitly selected. As a result, leptons refer to light leptons unless otherwise stated.
This paper explores the potential for BSM physics discovery in a model-agnostic way in final states with \llll or \lllll, while demonstrating sensitivity to specific benchmark models.

In high energy physics, anomaly detection (AD) is used as a technique to boost the sensitivity to BSM physics in a model-agnostic way. Both the ATLAS and CMS experiments have published searches involving AD techniques, ranging from weakly supervised searches~\cite{HDBS-2018-59, HMBS-2024-34}, semi-supervised searches~\cite{CMS-EXO-22-026, EXOT-2021-19} and fully unsupervised searches~\cite{EXOT-2022-07,HDBS-2019-23}. The majority of AD searches are performed in jet-focused final states. In this work, the first ever AD search is performed in multilepton final states. This analysis focuses on the branch of AD named outlier detection, where the type of anomalous events targeted are out-of-distribution data (excesses in tails) as opposed to over-densities such as bumps. Using generative models to explicitly calculate the probability density of events originating from SM processes, the sensitivity to rare excesses (SM or BSM) is boosted in a model-agnostic (denoted model-independent in the remainder) way. Due to the complex, high-dimensional input feature space used, interesting anomaly-enriched regions are constructed from the probability density that could have previously been overlooked.
%
%
%

Two separate searches are performed and referred to as `model-independent' and `model-dependent'. Both searches analyse the same pre-selected events but categorise them differently according to the multiplicity and flavour of various objects and the AD score. The model-independent search assumes BSM signals populating one signal region at a time, whereas the model-dependent search is designed to assess the sensitivity to specific BSM models by fitting the AD score in all regions in a simultaneous fit.

In order to assess the sensitivity of this search to BSM physics, a range of benchmark BSM scenarios are simulated. This not only demonstrates the generality of model-independent searches but allows comparison to the sensitivity of dedicated searches.
The benchmark models tested target the production of high-mass exotic particles that decay to final states with a high lepton multiplicity.
The reporting of model-independent limits allows future reinterpretations, and can be used as a benchmark to assess the viability of new BSM models in multilepton final states.

\begin{figure}[!h]
\centering
\subfloat[]{\includegraphics[width=0.33\textwidth]{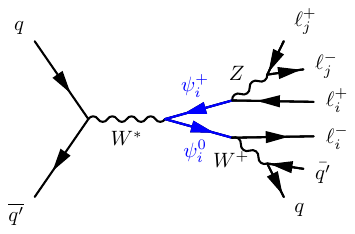}}
\subfloat[]{\includegraphics[width=0.33\textwidth]{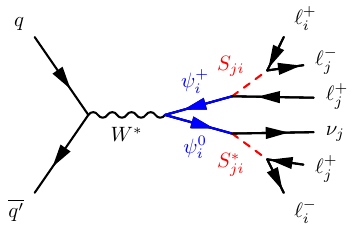}}
\subfloat[]{\includegraphics[width=0.33\textwidth]{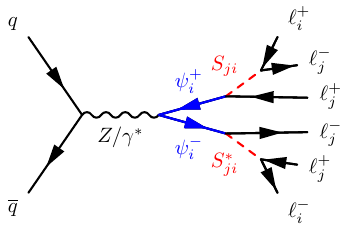}}
\caption{Example Feynman diagram for the pair production of VLL, denoted by $\psi_i$, in the flavourful model, with the VLLs produced via a (a, b) $W^*$ boson and (c) a $Z$ boson, with the decay of the VLLs into SM leptons and (a) two SM vector bosons, and (b, c) two scalars $S_{ji}$. Considering the decays of (a) $W^+ \to q \bar{q'}$ and $Z \to \ell_j^+ \ell_j^-$, and (b, c) $S^{(*)}_{ji} \to \ell_j^{-(+)} \ell_i^{+(-)}$, these processes result in multilepton final states. The subscripts $i$ and $j$ refer to the lepton generation, with allowed values $i=1,2$ and $j=1,2,3$.}
\label{fig:FlVllDiagrams}
\end{figure}

The benchmark models used for model-dependent interpretation in this paper include the production of vector-like leptons (VLLs) or of supersymmetric particles and are described in the following. Vector-like leptons are hypothetical spin-1/2 particles that are part of one of the simplest viable extensions to the SM at the electroweak scale~\cite{delAguila:1982fs,PhysRevD.33.2632,Fishbane:1987tx,Montvay:1988av,Fujikawa:1994we,DELAGUILA19901}. This analysis considers two simple minimal cases of VLLs as defined in Refs.~\cite{Kumar:2015tna,Bhattiprolu:2019vdu}, namely an ``SU(2) Singlet VLL'' and an ``SU(2) Doublet VLL'' model, with mixings to the SM electron or muon. VLLs are predominantly produced in pairs via the electroweak interaction.
In the SU(2) Doublet VLL scenario, the charged VLL decay modes are to an electron or muon, and a $Z$ or SM Higgs boson. In contrast, the neutral VLL decay mode is to an electron or muon and a $W$ boson with 100\% branching ratio. In the SU(2) Singlet VLL scenario, only the charged VLL is present and its decay modes are to a neutrino and a $W$ boson, an electron or muon and a $Z$ boson, or an electron or muon and a SM Higgs boson, with branching ratios asymptotically reaching 50\%, 25\%, and 25\%, respectively. The corresponding Feynman diagrams can be found in Ref.~\cite{EXOT-2021-31}.
VLL models are also considered with the inclusion of a new complex leptophilic scalar $S_{ji}$ which can decay through fermion mixing to leptons of generation $i$ and $j$, opening new decay modes of the VLLs, $\psi_i$, to this scalar, $\psi_i^{-} \to S_{ji}^{*} \ell_j^{-}$ or $\psi_i^{0} \to S_{ji}^{*} \nu_j$, and their conjugates~\cite{Bissmann2021Feb}. Once the VLL decay to the $S_{ji}$ scalar is open, the corresponding branching ratio is nearly 100\%.
The decay mode $S^{*}_{ji} \to \ell_j^{+} \ell_i^{-}$ with $i=1,2$ and $j=1,2,3$, and its conjugate, are considered, with equal branching ratios to the three $j$ lepton flavours for a given $i$ lepton flavour. This can result in final states with large lepton multiplicities (up to six leptons), and allow for lepton flavour violating decays.
Figure~\ref{fig:FlVllDiagrams} shows representative Feynman diagrams for this model, referred to as `flavourful VLL'.
Finally, simplified R-parity violating (RPV) SUSY signals are considered, where the lightest SUSY particle is a bino-like neutralino ($\tilde{\chi}_{1}^{0}$), and decays as: $\tilde{\chi}_{1}^{0} \rightarrow \ell^{\pm}_{k}\ell^{\mp}_{i/j}\nu_{j/i}$. Every pair production event of wino-like charginos and neutralinos  $\tilde{\chi}_{1}^{\pm} \tilde{\chi}_{1}^{\mp} \rightarrow \tilde{\chi}_{1}^{0} W^{\pm} \tilde{\chi}_{1}^{0} W^{\mp}$ and $\tilde{\chi}_{1}^{\pm} \tilde{\chi}_{2}^{0} \rightarrow \tilde{\chi}_{1}^{0} W^{\pm} \tilde{\chi}_{1}^{0} Z/h$ results in at least four lepton final states. In this paper, only light leptons are considered and thus the $\lambda_{ijk} = \lambda_{12k}$ (with $k=1,2$) coupling scenario is used as benchmark model.
The corresponding Feynman diagrams can be found in Ref.~\cite{SUSY-2018-02}.
A variation on the RPV SUSY model is also considered, where smuons are pair produced and decay to a bino-like neutralino, which in turn decays via the RPV coupling $\lambda_{i33}'$ (with $i=2,3$). This allows the bino-like neutralino to decay to a lepton and third-generation quarks. This model can result in final states with high lepton multiplicity and high $b$-jet multiplicity. The corresponding Feynman diagrams can be found in Ref.~\cite{HDBS-2020-03}.

%
%
%
%
%
%
%


%

%
\section{ATLAS detector}
%

%
\newcommand{\AtlasCoordFootnote}{%
ATLAS uses a right-handed coordinate system with its origin at the nominal interaction point (IP)
in the centre of the detector and the \(z\)-axis along the beam pipe.
The \(x\)-axis points from the IP to the centre of the LHC ring,
and the \(y\)-axis points upwards.
Polar coordinates \((r,\phi)\) are used in the transverse plane,
\(\phi\) being the azimuthal angle around the \(z\)-axis.
The pseudorapidity is defined in terms of the polar angle \(\theta\) as \(\eta = -\ln \tan(\theta/2)\) and is equal to the rapidity
$ y = \frac{1}{2} \ln \left( \frac{E + p_z}{E - p_z} \right) $ in the relativistic limit.
Angular distance is measured in units of \(\Delta R \equiv \sqrt{(\Delta y)^{2} + (\Delta\phi)^{2}}\).}

The ATLAS detector~\cite{PERF-2007-01} at the LHC covers nearly the entire solid angle around the collision point.\footnote{\AtlasCoordFootnote}
It consists of an inner tracking detector surrounded by a thin superconducting solenoid, electromagnetic and hadronic calorimeters,
and a muon spectrometer incorporating three large superconducting air-core toroidal magnets.

The inner-detector system (ID) is immersed in a \qty{2}{\tesla} axial magnetic field
and provides charged-particle tracking in the range of \(|\eta| < 2.5\).
The high-granularity silicon pixel detector covers the vertex region and typically provides four measurements per track,
the first hit generally being in the insertable B-layer (IBL) installed before Run~2~\cite{ATLAS-TDR-19,PIX-2018-001}.
It is followed by the SemiConductor Tracker (SCT), which usually provides eight measurements per track.
These silicon detectors are complemented by the transition radiation tracker (TRT),
which enables radially extended track reconstruction up to \(|\eta| = 2.0\).
The TRT also provides electron identification information
based on the fraction of hits (typically 30 in total) above a higher energy-deposit threshold corresponding to transition radiation.

The calorimeter system covers the pseudorapidity range \(|\eta| < 4.9\).
Within the region \(|\eta|< 3.2\), electromagnetic calorimetry is provided by barrel and
endcap high-granularity lead/liquid-argon (LAr) calorimeters,
with an additional thin LAr presampler covering \(|\eta| < 1.8\)
to correct for energy loss in material upstream of the calorimeters.
Hadronic calorimetry is provided by the steel/scintillator-tile calorimeter,
segmented into three barrel structures within \(|\eta| < 1.7\), and two copper/LAr hadronic endcap calorimeters.
The solid angle coverage is completed with forward copper/LAr and tungsten/LAr calorimeter modules
optimised for electromagnetic and hadronic energy measurements, respectively.

The muon spectrometer (MS) comprises separate trigger and
high-precision tracking chambers measuring the deflection of muons in a magnetic field generated by the superconducting air-core toroidal magnets.
The field integral of the toroids ranges between \num{2.0} and \qty{6.0}{\tesla\metre}
across most of the detector.
Three layers of precision chambers, each consisting of layers of monitored drift tubes, cover the region \(|\eta| < 2.7\),
complemented by cathode-strip chambers in the forward region, where the background is highest.
The muon trigger system covers the range \(|\eta| < 2.4\) with resistive-plate chambers in the barrel, and thin-gap chambers in the endcap regions.

The luminosity is measured mainly by the LUCID--2~\cite{LUCID2} detector that records Cherenkov light produced in the quartz windows of photomultipliers located close to the beampipe.

Events were selected by the first-level trigger system implemented in custom hardware,
followed by selections made by algorithms implemented in software in the high-level trigger~\cite{TRIG-2016-01}.
The first-level trigger accepted events from the \qty{40}{\MHz} bunch crossings at a rate close to \qty{100}{\kHz},
which the high-level trigger further reduced to record complete events to disk at about \qty{1.25}{\kHz}.

A software suite~\cite{SOFT-2022-02} is used in data simulation, in the reconstruction
and analysis of real and simulated data, in detector operations, and in the trigger and data acquisition
systems of the experiment.


%

%
\section{Data and simulated event samples}
\label{sec:datamc}

This analysis uses data from $pp$ collisions at $\sqrt{s}=13~\tev$ collected by the ATLAS experiment from 2015 to 2018. After the application of data-quality requirements~\cite{DAPR-2018-01} to ensure that all parts of the detector are operational during data-taking, the data sample corresponds to an integrated luminosity of \lumival.
The number of additional $pp$ interactions per bunch crossing (pile-up) in this sample ranges from about 8 to 70, with an average of 34.
The trigger requirements are discussed in Section~\ref{sec:events}.

Monte Carlo (MC) simulation samples are produced for the different signal and background processes.
Processes contributing to the background with at least one non-prompt lepton or one electron from conversion in the \llll final state are mainly \ttbar, \Zjets, \ttW and \WZ. The main background contributions with prompt leptons are from the \ttll, \ZZ, and \ttH processes. In the \lllll final state, \ttll and \ZZ processes contribute to the background with at least one non-prompt lepton or electron from photon conversion.

All samples showered with \textsc{Pythia} use the A14 set of tuned parameters~\cite{ATL-PHYS-PUB-2014-021} (referred to as `tune'), whereas those showered with \textsc{Herwig} use the H7-UE tune~\cite{Bellm:2015jjp}.
In all samples simulated with \Sherpa~\cite{Gleisberg:2008ta}, the matrix elements (MEs) are calculated with the Comix~\cite{Gleisberg:2008fv} and \OPENLOOPS~\cite{Buccioni:2019sur,Cascioli:2011va,Denner:2016kdg} libraries. They are matched with the \SHERPA parton shower (PS)~\cite{Schumann:2007mg} using the \MEPSatNLO prescription~\cite{Hoeche:2011fd,Hoeche:2012yf,Catani:2001cc,Hoeche:2009rj} with the set of tuned parameters developed by the \SHERPA authors.
Pile-up is modelled using events from minimum-bias interactions generated with \PYTHIA[8.186]~\cite{Sjostrand:2007gs} with the A3 tune~\cite{ATL-PHYS-PUB-2016-017}, and overlaid onto the simulated hard-scatter events according to the luminosity profile of the recorded data~\cite{SIMU-2020-01}.
All samples include leading-logarithm photon emission, either modelled by the PS generator or by \textsc{Photos}~\cite{Golonka:2005pn}.
The generated events are processed through either a full simulation of the ATLAS detector geometry and response using \textsc{Geant4}~\cite{Agostinelli:2002hh}, or a faster simulation where the full \textsc{Geant4} simulation of the calorimeter response is replaced by a detailed parameterisation of the shower shapes (ATLAS Fast Simulation)~\cite{SOFT-2010-01}. Both types of simulated events are processed through the same reconstruction software used for the $pp$ collision data.
Corrections are applied to the simulated events so that the particle candidates' selection efficiencies, energy scales and energy resolutions match those determined from data control samples.
Background events from $t\bar{t}Z/\gamma^\ast$ production were simulated using the \MGNLO[2.8.1]~\cite{Alwall:2014hca} generator at next-to-leading-order (NLO) in \alphas with the \NNPDF[3.0nlo]~\cite{Ball:2014uwa} PDF set.
The functional form of the renormalisation and factorisation scales (\muR, \muF) was set to the default scale $0.5 \times \sum_i \sqrt{m^2_i+p^{2}_{\textrm{T},i}}$, where the sum runs over all the particles generated from the ME calculation.
The invariant mass of the lepton pair, $\mll$, in the \ttll sample is set to be greater than $1~\gev$.
Top quarks were decayed at leading-order (LO) using \MADSPIN~\cite{Frixione:2007zp,Artoisenet:2012st} to preserve all spin correlations.
Showering and hadronisation were performed using \PYTHIA[8.244]~\cite{Sjostrand:2014zea} with the A14 tune, and the \NNPDF[2.3lo]~\cite{Ball:2012cx} PDF set with $\alphas = 0.130$. The decays of bottom and charm hadrons were performed by \EVTGEN[1.7.0]~\cite{Lange:2001uf}.
Diboson ($\VV$) background processes are simulated with \Sherpa[2.2.2]~\cite{Bothmann:2019yzt} and include $\WZ$, $\ZZ$, and $\WW$ processes.
The ME is calculated with NLO accuracy in QCD for up to one additional parton and at LO accuracy for up to three additional partons.
The \NNPDF[3.0nnlo] set of PDFs is used.
The simulation includes off-shell effects and Higgs boson contributions, where appropriate.
The invariant mass of any pair of same-flavour opposite-sign (SFOS) leptons is required to be $\mll > 4~\gev$.
Samples for the loop-induced processes $gg \to VV$ are simulated using LO-accurate MEs for up to one additional parton emission.
The triboson ($VVV$) background processes are also simulated with \Sherpa[2.2.2] and using the \NNPDF[3.0nnlo] set of PDFs, with the ME calculated with NLO accuracy in QCD for the inclusive process and LO accuracy for up to two additional partons.
Samples for $\tth$, and single top production are simulated using the NLO generator \POWHEGBOX[v2]~\cite{Frixione:2007nw,Nason:2004rx,Frixione:2007vw,Alioli:2010xd,Re:2010bp,Frederix:2012dh,Alioli:2009je} and interfaced with \PYTHIA[8] for the PS and fragmentation. These samples used the \NNPDF[3.0nlo] PDF set. The decays of bottom and charm hadrons are performed by \EVTGEN[1.6.0]. The production of a top quark in association with a \(W\) boson (\(tW\)) is modelled using the five-flavour scheme. The diagram removal scheme~\cite{Frixione:2008yi} is used to remove interference and overlap with \ttbar production. Single-top \(s\)- and \(t\)-channel production is modelled using the five- and four-flavour schemes, respectively.

The production of \ttbar events is modelled using the \POWHEGBOX[v2] generator at NLO with the \NNPDF[3.0nlo] PDF set. The events are interfaced to \PYTHIA[8.230] to model the PS, hadronisation, and underlying event, using the \NNPDF[2.3lo] set of PDFs. The decays of bottom and charm hadrons are performed by \EVTGEN[1.6.0]. The \ttbar process is modelled with the \hdamp parameter~\footnote{The \hdamp parameter is a resummation damping factor and one of the parameters that controls the matching of \POWHEG MEs to the PS, thus effectively regulating the high-\pT radiation against which the hard-process system recoils.} set to 1.5\,$m_t$~\cite{ATL-PHYS-PUB-2016-020}. The \ttbar sample is normalised to the \xs prediction at next-to-next-to-leading-order (NNLO) in QCD including the resummation of next-to-next-to-leading logarithmic (NNLL) soft-gluon terms calculated using \TOPpp[2.0]~\cite{Beneke:2011mq,Cacciari:2011hy,Baernreuther:2012ws,Czakon:2012zr,Czakon:2012pz,Czakon:2013goa,Czakon:2011xx}. This \xs is \(\sigma(\ttbar)_{\text{NNLO+NNLL}} = 832 \pm 51\,\si{\pb}\).

The $Z/\gamma^* \to  \ell \ell$ process (with $\ell=e,\mu,\tau$) is simulated with \Sherpa[2.2.11] using the \NNPDF[3.0nnlo] PDF set.
Processes with up to two coloured partons are modelled at NLO in the strong coupling, while processes with up to five additional partons are modelled at LO accuracy.
The $Z$+jets sample is normalised to the theoretical cross-section calculated at NLO accuracy in QCD~\cite{Catani:2009sm}.

The \ttW background sample is simulated using \MGNLO[2.6.7] with NLO multileg merging using the FxFx algorithm~\cite{Frederix:2012ps}, where the MEs are calculated for up to one additional partons at NLO in QCD, and up to two partons at LO in QCD\@. The events are interfaced with \PYTHIA[8.244]. The \NNPDF[3.0nlo] and \NNPDF[2.3lo] PDF sets are used for the ME and PS, respectively. The merging scale parameter is used in the matching of a ME with the PS and is set to 30 GeV.
In addition to this \ttW prediction at NLO in QCD, higher-order corrections related to electroweak (EW) contributions are also included. First, event-by-event correction factors are applied that provide virtual NLO EW corrections of the order $\alpha^{2}\alphas^{2}$ derived using the formalism described in Ref.~\cite{Kallweit:2015dum} along with LO corrections of order $\alpha^3$. Second, real emission contributions from the sub-leading EW corrections at order $\alpha^{3}\alphas$\,\cite{Frederix:2017wme} are simulated with \MGNLO[2.6.7] produced at LO in QCD and included as a separate sample.
A dedicated $\ttbar$ sample with rare $t \to Wb \gamma^*(\to \ell^{+}\ell^{-})$ radiative decays, $\ttbar \to W^+bW^-\bar{b}\ell^{+}\ell^{-}$, is simulated using a ME calculated at LO in QCD and requiring $\mll > 1~\gev$. In this sample the photon can be radiated from the top quark, the $W$ boson, or the $b$-quark. Both the \ttll and $\ttbar \to W^+bW^-\bar{b}\ell^{+}\ell^{-}$ samples are combined and together form the `$\ttdy$' sample.
The contribution from internal photon conversions ($\gamma^* \to \ell^{+}\ell^{-}$) with $\mll < 1~\gev$ is modelled by QED multi-photon radiation via the PS in an inclusive $\ttbar$ sample. The contribution from muons from internal conversions for $\mll < 1~\gev$ is negligible in this analysis.
Dedicated $Z$+jets samples are generated with \POWHEGBOX and interfaced with \PYTHIA[8] for the PS and fragmentation. These samples are used to model the data in a control region enriched in material and internal conversion electrons, as explained in Section~\ref{sec:events}.

The production of \tttt events was modelled using the \textsc{Madgraph5\_aMC@NLO}~v2.6.2 generator that provides MEs at NLO in QCD with the \NNPDF[3.1nlo]~\cite{NNPDF:2017mvq} PDF set. The functional form of the renormalisation and factorisation scales are set to $\mu_r = \mu_f = m_\mathrm{T} / 4$, where $m_\mathrm{T}$ is defined as the scalar sum of the transverse masses $\sqrt{m^2 + p_\mathrm{T}^2}$ of the particles generated from the ME calculation. The events are interfaced with \textsc{Pythia}~8.230 for the PS and hadronisation, using the \NNPDF[2.3lo] PDF set.
The production of \tttt\ events is normalised to a cross section of 12 fb computed at NLO in QCD including EW corrections~\cite{Frederix:2017wme}.
Other minor background contributions are included in the analysis: \tZ, \tWZ, \Wjets and $VH$ processes are normalised to their NLO cross-section, whereas $t\bar t t$, $t\bar t W^+ W^-$, $t \bar{t} ZZ$, $t \bar{t} HH$, and $t \bar{t} WH$ are normalised to their LO cross-section.

Signal samples for \vlle and \vllmu from $SU(2)$ singlet (\vlles, \vllmus) and doublet (\vlled, \vllmud) models, as well as the flavourful \vlle and \vllmu processes, are simulated using \MGNLO[2.9.5] at NLO in QCD with the \NNPDF[3.0nlo] PDF set and \PYTHIA[8.245], and processed through the ATLAS Fast Simulation.
The NLO cross-section obtained from \textsc{Madgraph} is used for the normalisation of the signals.
The RPV SUSY wino-like chargino and neutralino signal processes are simulated using \MGNLO[2.2.2] + \PYTHIA[8.230] at LO in QCD with up to two extra partons.
Jet--parton matching followed the CKKW-L prescription~\cite{Lonnblad:2012ix}, with a matching scale set to one quarter of the mass of the pair-produced SUSY particles.
The RPV SUSY smuon signal samples were generated with \MGNLO[2.9.3] with up to two extra jets at LO in QCD and \PYTHIA[8.245]. The matching scale is set at 1/4 of the mass of the SUSY particle being produced.
Cross-sections for the RPV SUSY signals were calculated to NLO in the strong coupling constant, adding the resummation of soft gluon emission at next-to-leading-logarithm accuracy~\cite{Beenakker:1996ch,Kulesza:2008jb,Kulesza:2009kq,Beenakker:2009ha,Beenakker:2011fu,Fuks:2012qx,Fuks:2013vua,Fuks:2013lya}.

\section{Event reconstruction}
\label{sec:objects}

Events containing light leptons, and with varying multiplicity of jets with and without $b$-hadrons, and missing transverse momentum content are selected. The corresponding object reconstruction is detailed below. Candidate events are required to have at least one $pp$ interaction vertex. Interaction vertices are reconstructed from at least two tracks with transverse momentum $\pt$ larger than $500~\mev$ that are consistent with originating from the beam collision region in the $x$--$y$ plane.
In events with multiple vertices, the primary vertex is defined as the one with the highest scalar sum of the squared transverse momenta of the associated tracks~\cite{ATL-PHYS-PUB-2015-026}.

Candidate events are selected with a different trigger strategy depending on the lepton multiplicity of the event: single-lepton triggers are used to select events in \llSS and \lll regions, and dilepton triggers are used in \llll and \lllll regions, requiring the electrons or muons to satisfy identification criteria similar to those used in the offline reconstruction and isolation requirements~\cite{TRIG-2018-01,TRIG-2018-05}.
Single-electron triggers require a minimum $\pt$ threshold of 24 (26)~$\gev$ in the 2015 (2016, 2017 and 2018) data-taking period(s), while single-muon triggers have a lowest $\pt$ threshold of 20 (26)~$\gev$ in 2015 (2016--2018).
The dielectron triggers require two electrons with minimum $\pt$ thresholds ranging from 12~$\gev$ in 2015 to 24~$\gev$ in 2017--2018, whereas the dimuon triggers use asymmetric $\pt$ thresholds for leading (subleading) muons: 18 (8)~$\gev$ in 2015 and 22 (8)~$\gev$ in 2016--2018. Finally, an electron+muon trigger requires events to have an electron candidate with a 17~$\gev$ threshold and a muon candidate with a 14~$\gev$ threshold for all periods.
In the offline selection, leptons are required to match, with $\Delta R < 0.15$, the corresponding leptons reconstructed by the trigger and to have a $\pt$ exceeding the trigger $\pt$ threshold by at least 1~$\gev$.

Electron candidates are reconstructed from energy clusters in the electromagnetic calorimeter matched to a track in the ID~\cite{EGAM-2018-01}. They are required to satisfy $\pt>10$~GeV and $|\eta|<2.47$, excluding the transition region between the endcap and barrel calorimeters ($1.37<|\eta|<1.52$).
The \enquote{Loose} electron identification working point (WP) is used, based on a likelihood discriminant employing calorimeter, tracking and combined information that provide separation between electrons and jets.

The reconstruction of muon candidates is based on tracking information from the MS and the ID, as well as energy deposits in the calorimeter system~\cite{MUON-2018-03}. Muons are required to satisfy $\pt > 10~\gev$ and $|\eta| < 2.5$. The \enquote{Loose} muon identification WP is used.

Electron (muon) candidates are matched to the primary vertex by requiring that the significance of their transverse impact parameter, $d_0$ (defined as the distance of closest approach of the track to the beamline in the $x$--$y$~plane) satisfies $|d_0|/\sigma_{d_0}<5\,(3)$, where $\sigma_{d_0}$ is the measured uncertainty in $d_0$, and by requiring that their longitudinal impact parameter, $z_0$ (defined as the distance in $z$ between the primary vertex and the point on the track used to evaluate $d_0$) satisfies $|z_0 \sin\theta|<0.5$~mm.

Light lepton candidates are also required to be isolated in the tracker and in the calorimeter to further suppress leptons from heavy-flavour (HF) hadron decays, misidentified jets, or photon conversions (collectively referred to as \enquote{non-prompt leptons}). Electrons with an incorrect charge assignment (referred to as `QMisID') are further suppressed with a boosted decision tree (BDT) discriminant~\cite{PERF-2017-01} in \llll events with total lepton electric charge \qtwo and same-sign electric charge dilepton (\llSS) $\mu e$ events. Electrons are further separated into three classes, \enquote{material conversion}, \enquote{internal conversion}, and \enquote{non-conversion}, following the same procedure as described in Ref.~\cite{EXOT-2021-31}. Electrons are required to pass the \enquote{non-conversion} class in all regions except in the control region enriched in material and internal conversion electrons, as explained in Section~\ref{sec:events}.

For the control regions enriched in non-prompt leptons from the decay of hadrons that contain bottom- or charm-quarks (referred to as \enquote{HF non-prompt leptons}) as described in Section~\ref{sec:events}, leptons are required to fail a selection based on a WP of a BDT discriminant (referred to as the non-prompt-lepton BDT~\cite{HIGG-2017-02}). The BDT uses isolation and lifetime information about a track-jet that matches the selected electron or muon to discriminate between prompt and non-prompt leptons.

The constituents for jet reconstruction are identified by combining measurements from both the ID and the calorimeter using a particle flow (PFlow) algorithm~\cite{PERF-2015-09}. Jet candidates are reconstructed from these PFlow objects using the \antikt~algorithm~\cite{Cacciari:2008gp,Fastjet} with a radius parameter of $R=0.4$.
They are calibrated based on jet energy scale (JES) and resolution (JER), as derived from 13 TeV data and simulation~\cite{JETM-2018-05}, such that in simulation they have on average the same energy and momentum as matched particle level jets.
Only jet candidates with $\pT> 25$~\GeV\ and within $|\eta|<2.5$ are selected. To reduce the effect of \pileup, each jet with $\pT<60$~\GeV\ and $|\eta|<2.4$ is required to have an origin compatible with the primary vertex, as defined by the jet vertex tagger (JVT)~\cite{PERF-2014-03} criteria. A set of quality criteria is also applied to reject events containing at least one jet arising from non-collision sources or detector noise~\cite{ATLAS-CONF-2015-029}.

Jets containing $b$-hadrons are identified ($b$-tagged) via an algorithm~\cite{FTAG-2019-07} that uses a deep-learning neural network based on the distinctive features of $b$-hadron decays, primarily the impact parameters of tracks and the displaced vertices reconstructed in the ID\@. Additional input to this network is provided by discriminating variables constructed by a recurrent neural network, which exploits the spatial and kinematic correlations between tracks originating from the same $b$-hadron. A multivariate $b$-tagging discriminant value is calculated for each jet.
In this search, a jet is considered $b$-tagged if it passes the WP corresponding to 77\% average expected efficiency~\footnote{Efficiencies for tagging $b$-jets are determined for jets with $\pt >20~\gev$ and $|\eta|<2.5$ in simulated $\ttbar$ events.} to tag a $b$-quark jet in all regions except for the fake light-flavour electron control region where the 85\% b-tagging efficiency WP is used.
The corresponding light-jet\footnote{`Light jet' refers to a jet originating from the hadronisation of a light quark ($u$, $d$, $s$) or a gluon.} rejection factor~\footnote{The rejection factor is defined as the reciprocal of the efficiency.} is about 40 to 192, and the charm-jet ($c$-jet) rejection factor is about 3 to 6.
The $b$-tagging distribution obtained by ordering from higher to lower $b$-jet efficiency the resulting five exclusive bins from the four WPs corresponding to 85\%, 77\%, 70\%, or 60\% average expected efficiency to tag a $b$-quark jet, is referred to as pseudo-continuous $b$-tagging score. Each jet is assigned a pseudo-continuous $b$-tagging score that defines if a jet passes a given operating point but fails the adjacent tighter one. The sum of the pseudo-continuous $b$-tagging scores of all jets in the event is used as input to the AD discriminant described in Section~\ref{sec:events}.
Correction factors derived from dedicated calibration samples enriched in \bjet{}s, $c$-tagged jets, or light-tagged jets, are applied to the simulated event samples~\cite{FTAG-2018-01,FTAG-2020-08,FTAG-2019-02}.

To uniquely identify objects, a sequential \enquote{overlap removal} procedure is performed. If two electrons are separated by $\Delta R<0.1$, only the one with the higher \pt\ is kept. If an electron and a muon overlap within $\Delta R<0.1$, the muon is removed if it is reconstructed only from an ID track and calorimeter energy deposits consistent with a minimum-ionising particle (i.e.\ if it is \enquote{calo-tagged}), otherwise the electron is removed. If an electron and a selected jet are found within $\Delta R < 0.2$, the jet is removed. For each electron in the event a \pt-dependent variable-size cone of maximum size $\Delta R=0.4$ is defined. If a selected jet, surviving all previous overlap criteria, is found in this cone, the lepton is rejected. The same procedure is also applied between jets and muons, with the exception that, if a muon and a jet overlap with $\Delta R <$ 0.2, the jet is removed, unless the number of tracks in the jet is more than two.

The missing transverse momentum $\vec{p}_{\mathrm{T}}^{\mathrm{miss}}$ (with magnitude $\met$) is defined as the negative vector sum of the $\pt$ of all selected and calibrated objects in the event that satisfy the overlap removal procedure, and an additional term to account for the momenta of soft particles that are not associated with any of the selected objects~\cite{JETM-2020-03}.
This soft term is calculated from inner-detector tracks matched to the primary vertex, which makes it more resilient to contamination from \pileup interactions.


%

%
\section{Search region definition}
\label{sec:events}

%

%
Events containing at least four light leptons are selected and further divided into three orthogonal final states:
\begin{itemize}
\item \llll \qzero: four light leptons with a total charge of zero,
\item \llll \qtwo: four light leptons with a total charge of $\pm 2$, and
\item \lllll: five or more light leptons.
\end{itemize}
The total number of observed events in \llll \qzero, \llll \qtwo, and \lllll are 4097, 77, and 21, respectively.

Two searches are defined: one following a model-independent strategy consisting of `discovery regions', and the other designed to be sensitive to the benchmark models described in Section~\ref{sec:intro}, consisting of `benchmark regions'. The different event splitting in each of these regions is described below. However the events in the sum of all discovery regions are the same as the events in the sum of all benchmark regions.
Regions are defined through a trade-off between various considerations. First, the sensitivity to specific multilepton final states, covering possible combinations of number of leptons, lepton flavour, number of b-jets, number of SFOS lepton pairs, and those consistent with a $Z$ boson decay. Second, the granularity of the region splitting is limited by the available MC background simulated events to ensure a reliable background estimation in all regions. Third, the reduction of the look-elsewhere effect in the model-independent search is achieved with a smaller region multiplicity than in the model-dependent search.
In the \llll \qzero channel, discovery regions are defined by the number of SFOS lepton pairs, and by the number of SFOS lepton pairs with an invariant mass within 10~\gev of the nominal $Z$ boson mass, referred to as a $Z$ boson candidate, allowing the separation of on-shell $Z$ boson processes from off-shell and photon backgrounds.
The benchmark regions have the same categorisation of events as the discovery regions, with further splitting based on the $b$-tagged jet multiplicity (zero or $\geq1$ $b$-jets) and the number of electrons and muons that are not part of a $Z$ boson candidate ($N_{e}>N_{\mu}$, $N_{e}=N_{\mu}$ and $N_{\mu}>N_{e}$).
The $b$-jet-based classification improves the discrimination against top-quark processes. The main background contributions originate from \ZZ and \ttdy processes.

In the \llll \qtwo channel, where the main background processes contain at least one non-prompt lepton, the discovery region is not split. The benchmark regions are split into events with zero or $\geq1$ $b$-jets, and these categories are further divided into three sub-regions based on $N_{e}$ and $N_{\mu}$ as done in the \qzero channel. The $Z$ boson candidates are defined as in the \llll \qzero channel but without requiring the leptons to have opposite charge.

In the \lllll channel, the discovery regions are defined by the number of $Z$ boson candidates: 5$\ell$~0Z and 5$\ell$~$\geq$1Z. To define the benchmark regions, events are further split by the presence of $b$-jets (0 or $\geq1$ $b$-jets) and by lepton flavour when there are no $b$-tagged jets, due to the limited number of data events in $b$-tagged regions. The main background processes are \VVV, \ttH, and events with at least one non-prompt lepton.
Five control regions (CRs) orthogonal to the discovery/benchmark regions and among each other are defined to fit the normalisation of the leading backgrounds.
A region enriched in $\WZ$ and \ttZ is defined (`WZ/ttZ') by selecting events with three leptons, from which one SFOS lepton pair is required to be compatible with a $Z$ boson, $|\mllOSSF - m_Z| < 10~\GeV$. Events in the WZ/ttZ CR are also required to have at least one jet and zero $b$-jets satisfying the 77\% WP.
A region enriched in photon conversions (`Conversions') from $Z \rightarrow \mu \mu \gamma^{(*)}(\rightarrow ee)$ is defined by requiring three leptons (two muons and one electron), from which the electron must fulfil the material conversion or internal conversion candidate requirements. Events in the Conversion CR are also required to have no SFOS lepton pairs compatible with a $Z$ boson, the three lepton invariant mass fulfilling $|m_{3\ell} - m_Z| < 10~\GeV$, and zero $b$-jets satisfying the 77\% WP.
For the two CRs enriched in HF non-prompt leptons, events with two same-sign leptons are selected, requiring the leading lepton to additionally pass the non-prompt-lepton BDT WP, and the same-sign lepton pair to be $\mu e$ with the electron being the sub-leading lepton and passing the QMisID BDT requirement (CR enriched in non-prompt electrons from HF hadron decays, `HFe') or $\mu \mu$ (CR enriched in non-prompt muons from HF hadron decays, `HF$\mu$'). Additionally, at least two jets and exactly one $b$-jet satisfying the 77\% WP are required.
Finally, a CR enriched in non-prompt electrons from LF hadron decays (`LFe') consists of events with 3$\ell$, from which two same-flavour muons with an invariant mass consistent with that of a $Z$ boson plus one additional electron are required. In addition, the \met has to be lower than 20 GeV to reduce the $\WZ$ background
contamination and zero $b$-jets satisfying the 85\% WP are required.
The contamination in the control regions from the non-excluded signal benchmark models described in Section~\ref{sec:intro} is negligible.
%


%

%
\section{Anomaly detection}
\label{sec:anomaly}

%
An event-level AD technique is used based on normalising flows to enhance the sensitivity to BSM physics in a model-independent way. The AD technique is designed to be sensitive to out-of-distribution data (outlier detection), and it is thus sensitive to BSM physics populating low probability regions of phase space. Anomalies are defined with respect to a reference distribution, for which the simulated SM background is used.

To build an anomaly score that is sensitive to rare (kinematically extreme) events, the normalising flows are used to directly evaluate the probability density of the MC background. Normalising flows are an unsupervised machine learning model typically used to model complex distributions, allowing efficient and exact density estimation~\cite{rezende2016variationalinferencenormalizingflows}. To do so, flows employ a series of invertible bijective functions to transform from a closed-form latent distribution to a target distribution, allowing the explicit calculation of the probability density. In this work, a Real-valued Non-Volume Preserving flow (RealNVP) is used with the standard choice of a Normal distribution in the latent space~\cite{Dinh2016May}. In order to use the normalising flows in the context of AD, the flow is trained on the background MC prediction only, using the background probability density $p(x)$, where $x$ is the set of input variables, as the target distribution within the search regions.
The training is performed on MC samples rather than data due to the available statistics in multilepton final states and additional complexities with biasing when using data to train the normalising flows.
Once learned, the evaluated probability is then transformed to give the anomaly score according to:
\begin{equation}
s(x) = \frac{\log{p(x)} - \log{p_{\text{max}}}}{\log{p_{\text{min}}} - \log{p_{\text{max}}}},
\end{equation}

where $p_{\text{min}}$ and $p_{\text{max}}$ are the minimum and maximum probability densities, respectively, evaluated from the MC background dataset. The transformation is chosen such that the log probability is scaled to be within the range [0,1] with anomalous events at higher score values.

The choice of input variables used to calculate the background density is crucial for defining the space of BSM models that the AD is sensitive to. Physically motivated high-level variables are chosen which can provide sensitivity to a wide range of BSM models while also aiding in the characterisation of any excess that may be found.
Table~\ref{tab:InputVars} shows the training variables used in each search region. The pairing of leptons with the closest invariant mass to that of the $Z$ boson is denoted by the label $Z$ (``lepton pair closest to $Z$ boson''), and the remaining pair of leptons amongst the four leptons in the event (``lepton pair second-closest to $Z$ boson'') is denoted by the label $\ell\ell$.
In the \lllll region, a reduced number of input variables are used due to the relatively small sample size of the simulated events and the training is performed only on events with at least one $Z$ boson candidate. No AD is applied to \lllll 0Z region, where event counting is used instead due to limited data yields.

\begin{table}[!h]
\centering
\caption{A summary of the input variables used for training the normalising flows for AD in the \llll \qzero, \llll \qtwo, and $\lllll$ regions. Each variable is scaled to mean zero and unit variance before being used as training data.
The ``lepton pair closest to $Z$ boson'' is given by the SFOS lepton pair with invariant mass closest to 91.2 GeV, with the remaining pair of leptons amongst the four leptons in the event referred to as the ``lepton pair second-closest to $Z$ boson''.}
\begin{tabular}{cl}
\toprule
Training variable &  Description\\
\hline\hline
\multicolumn{2}{c}{Input to \llll \qzero, \llll \qtwo, and $\lllll$ 1Z} \\
\hline
\HTlep 		& Sum of transverse momenta of leptons \\
\HTjets 	& Sum of transverse momenta of jets \\
\met 		& Missing transverse energy\\
\njets 		& Number of jets \\
\ptZ 		& Transverse momentum of lepton pair closest to $Z$ boson \\
\hline
\multicolumn{2}{c}{Input to \llll \qzero and \llll \qtwo} \\
\hline
\ptll 		& Transverse momentum of lepton pair second-closest to $Z$ boson \\
\mZ 		& Invariant mass of lepton pair closest to $Z$ boson \\
\mll 		& Invariant mass of lepton pair second-closest to $Z$ boson \\
\mthreelh 	& Largest invariant mass of lepton pair closest to $Z$ boson and another lepton \\
\mthreell 	& Smallest invariant mass of lepton pair closest to $Z$ boson and another lepton  \\
\mfourl 	& Invariant mass of four-lepton system  \\
\mtfourlmet & Transverse mass of four leptons and \met \\
\mtzmet 	& Transverse mass of lepton pair closest to $Z$ boson and \met\\
\mtllmet	& Transverse mass of lepton pair second-closest to $Z$ boson and \met\\
\sumpcb 	& Sum of pseudo-continuous $b$-tagging score\\
\bottomrule
\end{tabular}
\vspace{0.5cm}
\label{tab:InputVars}
\end{table}

The same AD training is used for the model-dependent and independent analyses.
Normalising flows are trained in each of the discovery regions. In the \llll regions the flows are trained in two sub-regions, after splitting events according to the presence or absence of $b$-jets. The anomaly score distributions are then merged into one distribution for the model-independent analysis, while they are used separately for the model-dependent analysis. The \texttt{PyTorch} library was used to train the normalising flows used in this work, and the \texttt{normflows} package used to define the architecture of the flows~\cite{PyTorch,normflows_package}.
The flows are made up of eight affine coupling layers, each containing two hidden layers of fully connected neural networks (FCN), each with 64 nodes. Each FCN uses the \texttt{Tanh} activation function, and the parameters of the flows are initialised to zero. The Adam optimiser~\cite{kingma2017adam} is used within back-propagation, along with a small weight decay of $10^{-4}$ (an L2 regularisation on the model weights). Trainings are performed with a fixed learning rate of $10^{-4}$ or $10^{-5}$ depending on the available statistics in the region used, and with a fixed batch size of 512. The flows are trained using weighted MC events, with overall weights calculated through the combination of weights and scale factors as described in Section~\ref{sec:objects}. Weights are applied to the calculated loss per-event to produce the correctly weighted probability density. Events with negative weights are rare and removed from the training to improve stability.

Once the final anomaly score distributions are obtained, two different strategies are used to bin the distribution. For the model-independent search, background rejection points of 90\%, 99\% and 99.9\% are used to define overlapping discovery signal bins. Each discovery signal bin is required to have at least 0.1 expected background events and less than 20\% MC statistical uncertainty, or otherwise the bin is dropped. Discovery regions where the 90\% bin is dropped use a 50\% background rejection point, as long as the above criteria on background events and statistical uncertainty are fulfilled; otherwise, no AD is used in the corresponding region. The remaining low-anomaly score distribution is used as a region for measuring background normalisation effects. For the model-dependent search, the most sensitive bin is defined by tightening the selection on the anomaly score until both threshold criteria on the 0.1 background and 20\% MC statistical uncertainty are satisfied. From there, successive bins are defined such that the background yield increases in each bin by a factor of four until the entire distribution is binned. This results in distributions with higher granularity at high-anomaly score, and conversely low granularity at low-anomaly score.

\begin{figure}[tb]
\centering
\subfloat[]{\includegraphics[width=0.45\textwidth]{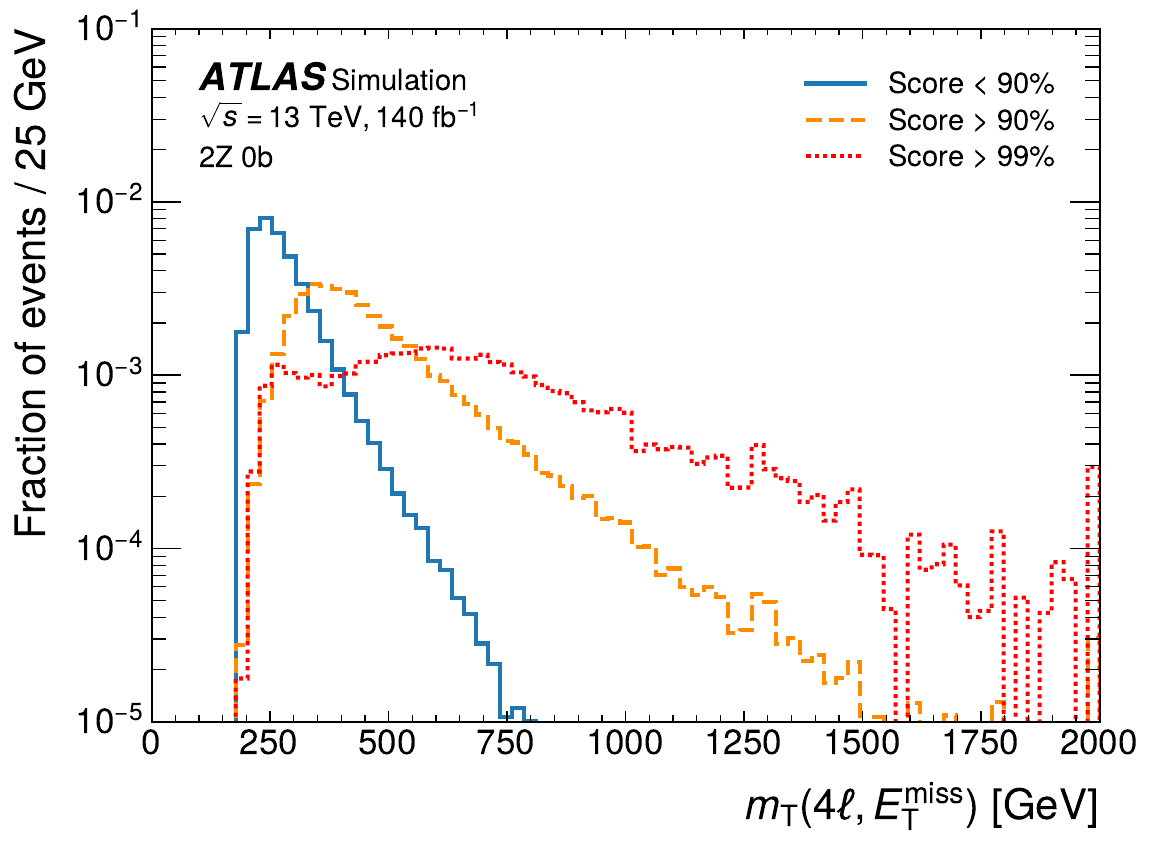}}
\subfloat[]{\includegraphics[width=0.45\textwidth]{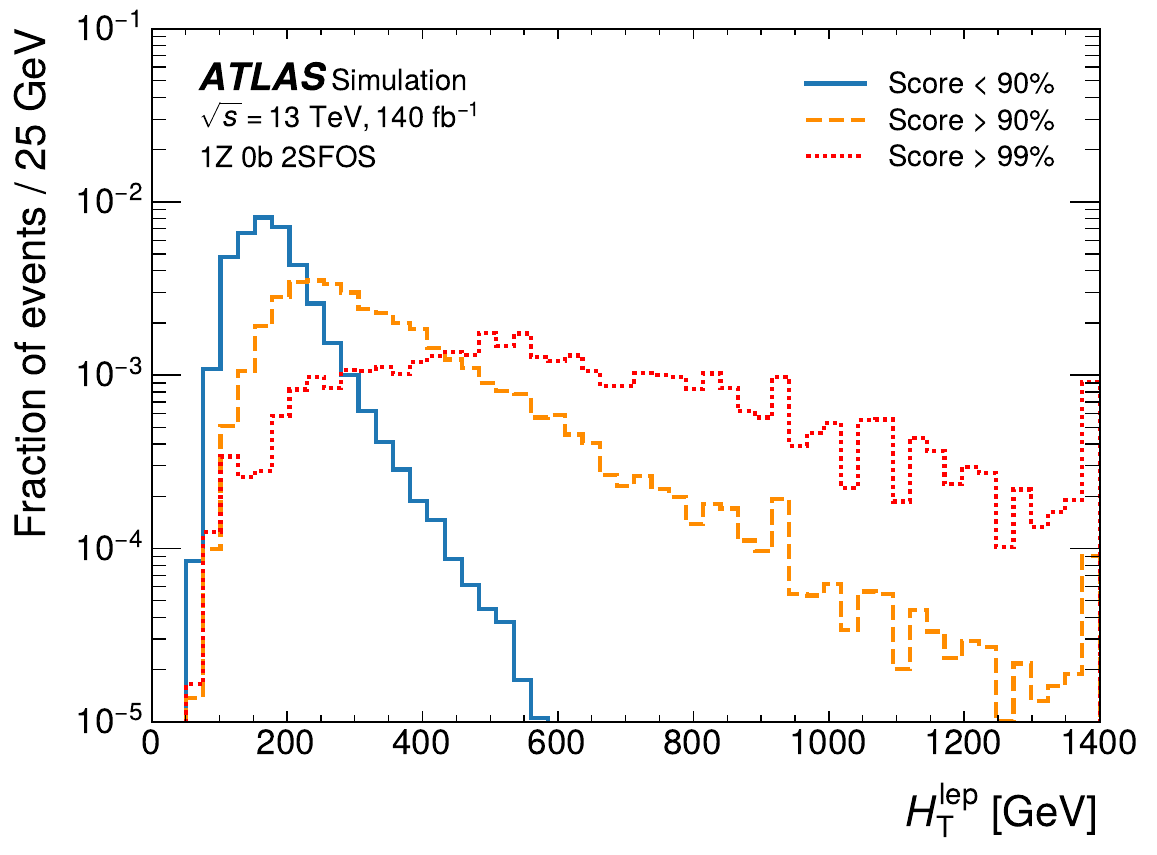}}
\caption{Comparison of background shapes for (a) \mtfourlmet in the 2Z 0b region and (b) \HTlep in the 1Z 0b 2SFOS region,
considering various requirements on the anomaly score corresponding to the specified background rejection cuts.
The last bin contains the overflow.}
\label{fig:sensitivity_separation}
\end{figure}

In order to quantify which input features the AD technique is most sensitive to, the distortion of the kinematic variables is quantified after applying a series of cuts to the anomaly score. For this purpose, the Energy Distance~\cite{EnergyDistance2} is calculated between each input variable after separating by low- and high-anomaly score. Figure~\ref{fig:sensitivity_separation} shows two example input variable distributions and the distortions after applying the specified background rejection cut to the anomaly score. The most impactful input variables are typically \mtfourlmet, \mll, \ptZ, \mfourl, and \HTlep.

\begin{figure}[tb]
\centering
\includegraphics[width=1\textwidth]{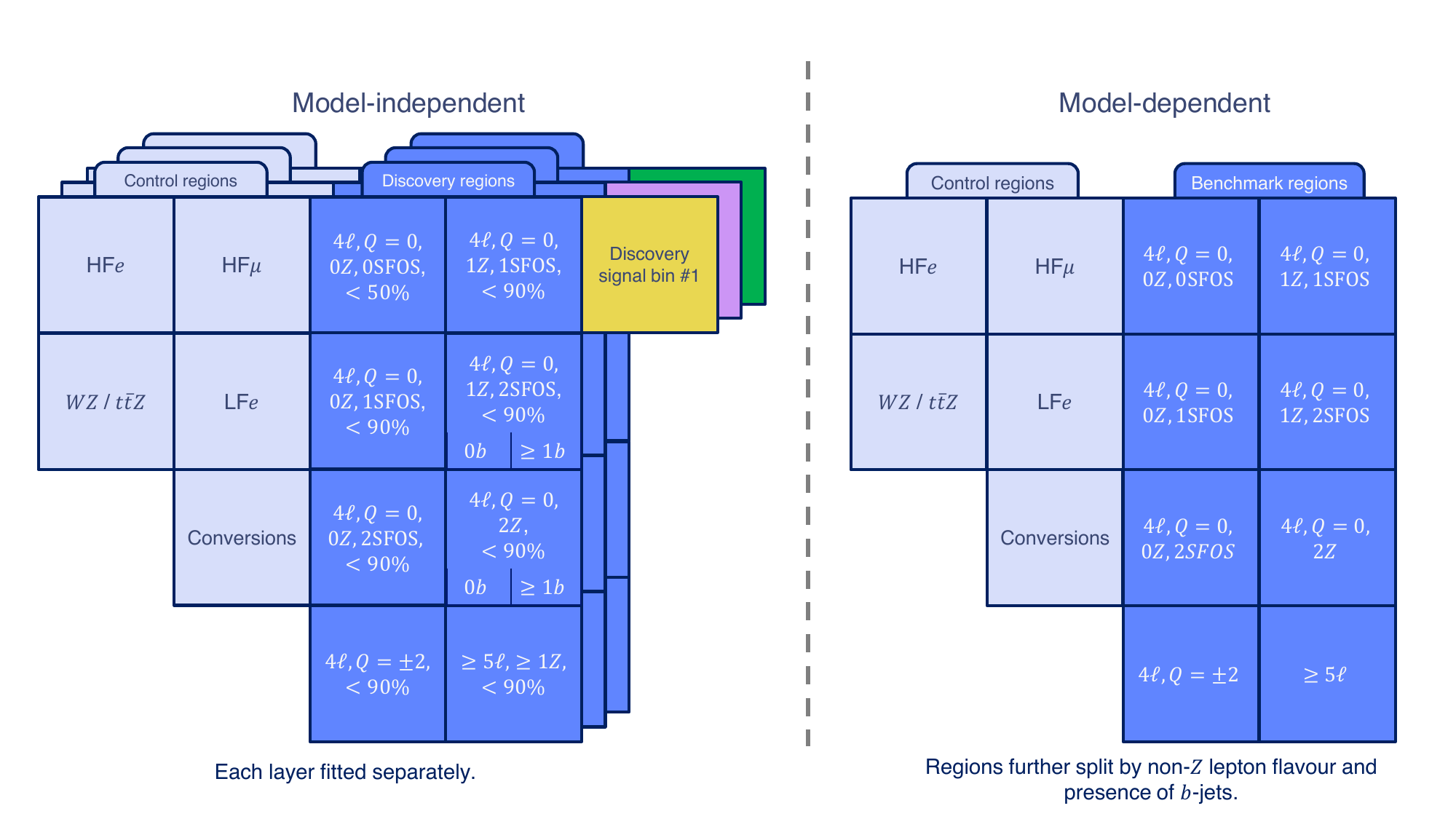}\\
\caption{Illustrative sketch of the control regions and (left) discovery or (right) benchmark regions in the (left) model-independent or (right) model-dependent search. For the model-independent scenario, the discovery regions correspond to low-anomaly score regions fitted together with a high-anomaly score discovery signal bin at a time. Each layer corresponds to the different fits with the same control regions and low-anomaly score regions, and varying signal bins. The <90\% (<50\%) label corresponds to the requirement on the anomaly score to be below the 90\% (50)\% background rejection point. For the model-dependent scenario, all benchmark regions are used simultaneously together with the control regions in a single profile likelihood fit to data, illustrated with one single layer.}
\label{fig:sketch_strategy}
\end{figure}

Two distinct fit setups, as defined in Section~\ref{sec:results}, are used to perform the model-independent and model-dependent searches, as illustrated schematically in Figure~\ref{fig:sketch_strategy}. Both searches share the same five control regions (`HFe', `HF$\mu$', `LFe',`Conversions',`WZ/ttZ'; coloured in light blue in Figure~\ref{fig:sketch_strategy}). The model-independent search includes ten additional low-anomaly score discovery regions (`Discovery regions'; coloured in darker blue in Figure~\ref{fig:sketch_strategy} (left)). These correspond to the pre-selected events categorised following the discovery region definitions in Section~\ref{sec:events}, not passing the discovery signal bin anomaly score cuts (denoted as ``<90\%'' or ``<50\%''). The low-anomaly score discovery regions \llll \qzero 1Z 2SFOS and \llll \qzero 2Z are further split in events with zero or at least one $b$-jet.
Each layer in Figure~\ref{fig:sketch_strategy} (left) represents the model-independent fit performed with the five control regions, the ten low-anomaly score discovery regions, and one discovery signal bin at a time. The discovery signal bin (the stack of differently coloured boxes in Figure~\ref{fig:sketch_strategy} (left)) is defined by the discovery region selection and the cut at the optimised background rejection value ``>99.9/99/90/50\%''. In total there are 16 discovery signal bins corresponding to 16 different fits. Fitting each discovery signal bin allows to probe BSM physics in different multilepton final states.
The model-dependent search includes the same pre-selected events as the model-independent search but additionally splitting them according to the flavour of the non-$Z$ leptons and the presence of 0 or $\geq1$ $b$-jets (`Benchmark regions'; coloured in darker blue in Figure~\ref{fig:sketch_strategy} (right)). A single fit per benchmark signal hypothesis is performed with the five control regions and 32 benchmark regions (and thus only one layer is displayed in Figure~\ref{fig:sketch_strategy} (right)), fitting the anomaly score distribution in all regions except in the \lllll 0Z regions, where the total event yield is used instead.

%

%
\section{Background estimation}
%

%

All background processes are estimated by using the simulation samples described in Section~\ref{sec:datamc}. Before the simultaneous fit to data, the event kinematics of the simulated $\ttbar$ and $\VV$ backgrounds require dedicated corrections derived from data control samples to better describe the data. During the simultaneous fit to data discussed in Section~\ref{sec:results}, the yields of \ZZ with additional light-flavour jets with zero, one or two on-shell $Z$ boson candidates ($\ZZLFzeroZ$, $\ZZLFoneZ$, $\ZZLFtwoZ$), \ZZ with additional heavy-flavour jets ($\ZZHF$), \WZ with additional light- and heavy-flavour jets ($\WZLF$, $\WZHF$), $\ttZ$, and non-prompt-lepton backgrounds, are adjusted via normalisation factors included in the likelihood fit, as described in Section~\ref{sec:results}.

The main background contributions with prompt leptons in the regions with $\geq 4 \ell$ originate from $\ZZ$, $\ttdy$, and $\VVV$ production. Smaller contributions originate from the following rare processes: $\ttH$, $\tttt$, $\tZ$, $\tW$, $\tWZ$, $\ttWW$, and $\ttt$ production.

\begin{figure}[t]
\centering
\subfloat[]{\includegraphics[width=0.32\textwidth]{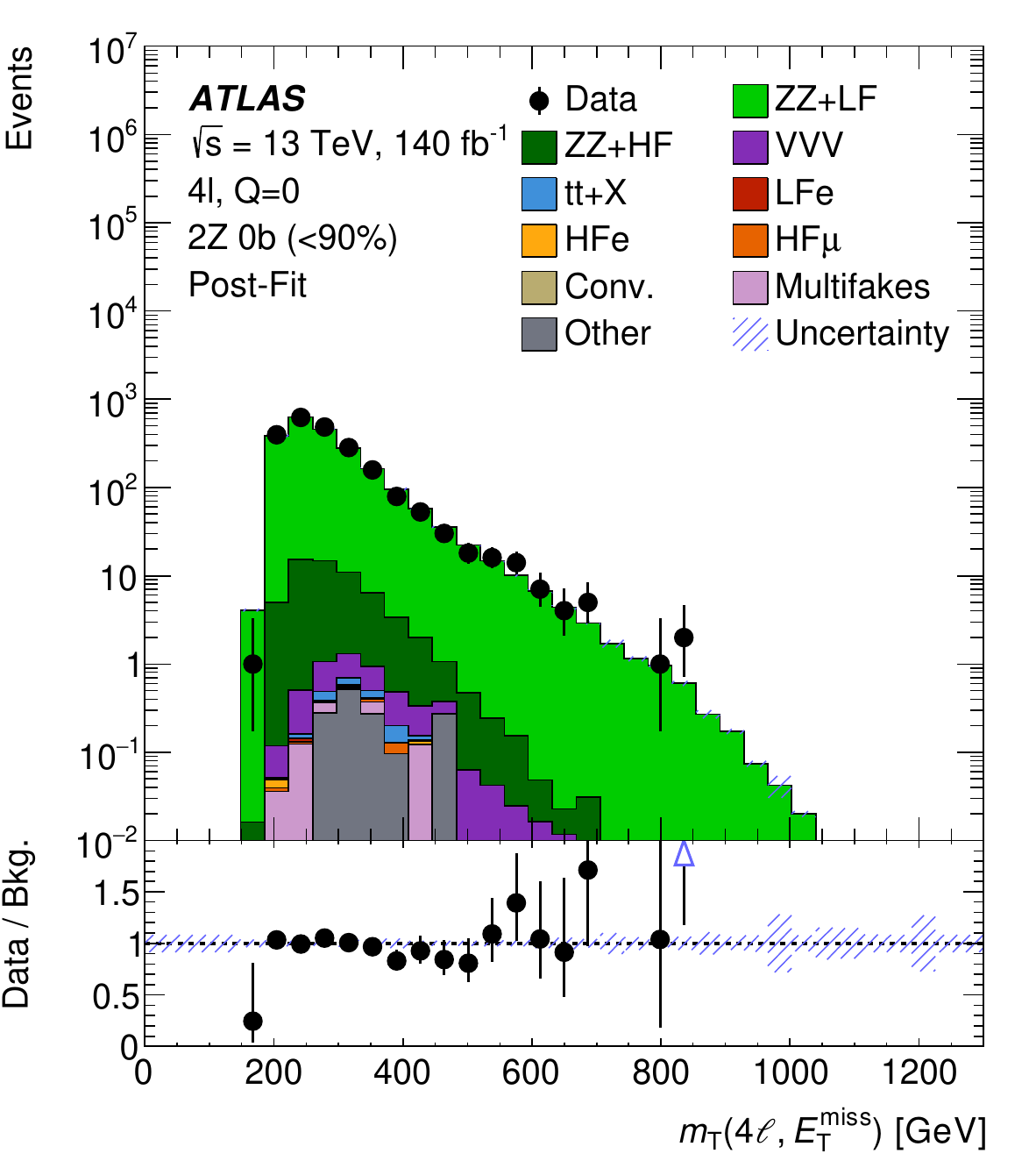}}
\subfloat[]{\includegraphics[width=0.32\textwidth]{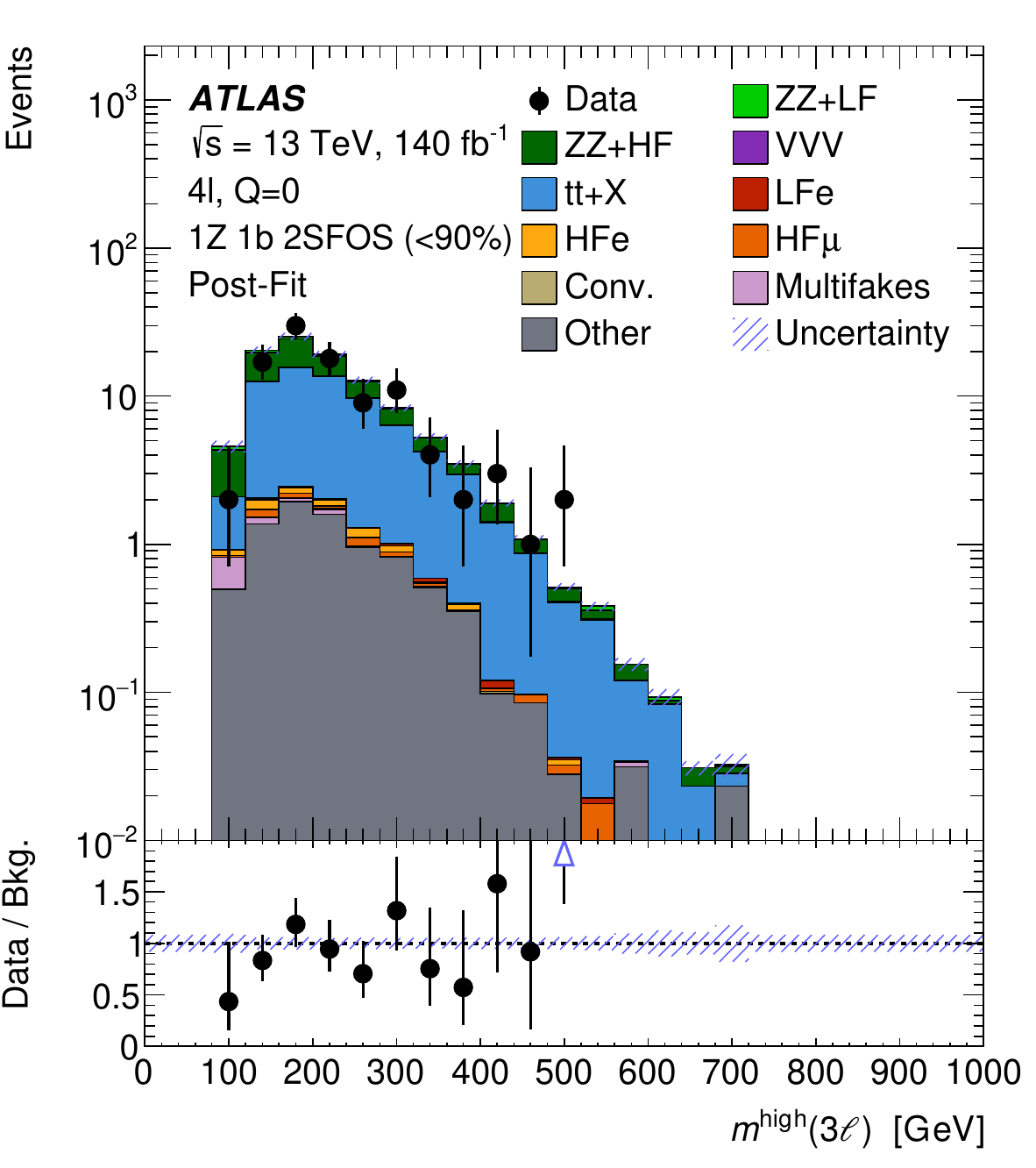}}
\subfloat[]{\includegraphics[width=0.32\textwidth]{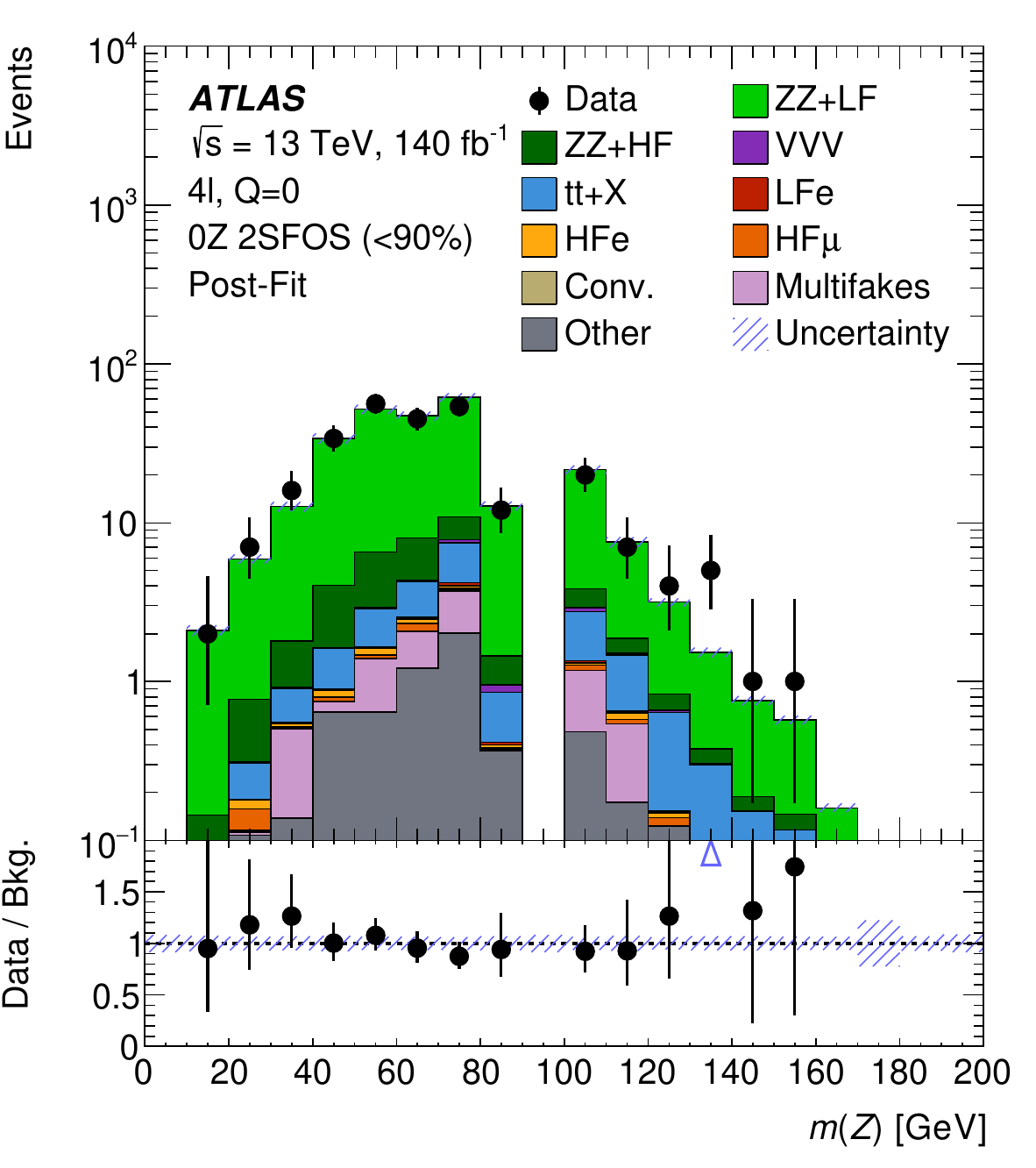}} \\
\subfloat[]{\includegraphics[width=0.32\textwidth]{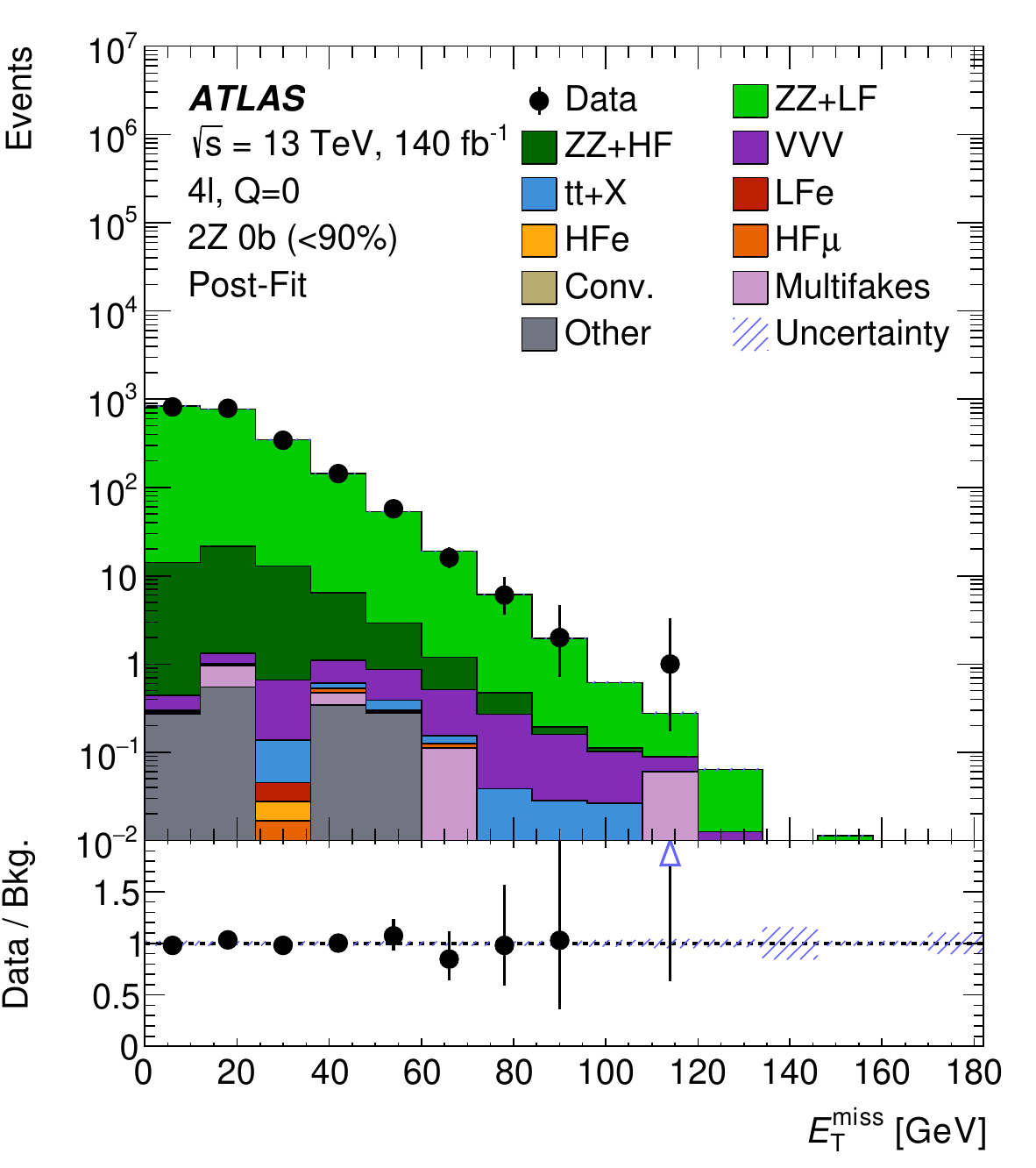}}
\subfloat[]{\includegraphics[width=0.32\textwidth]{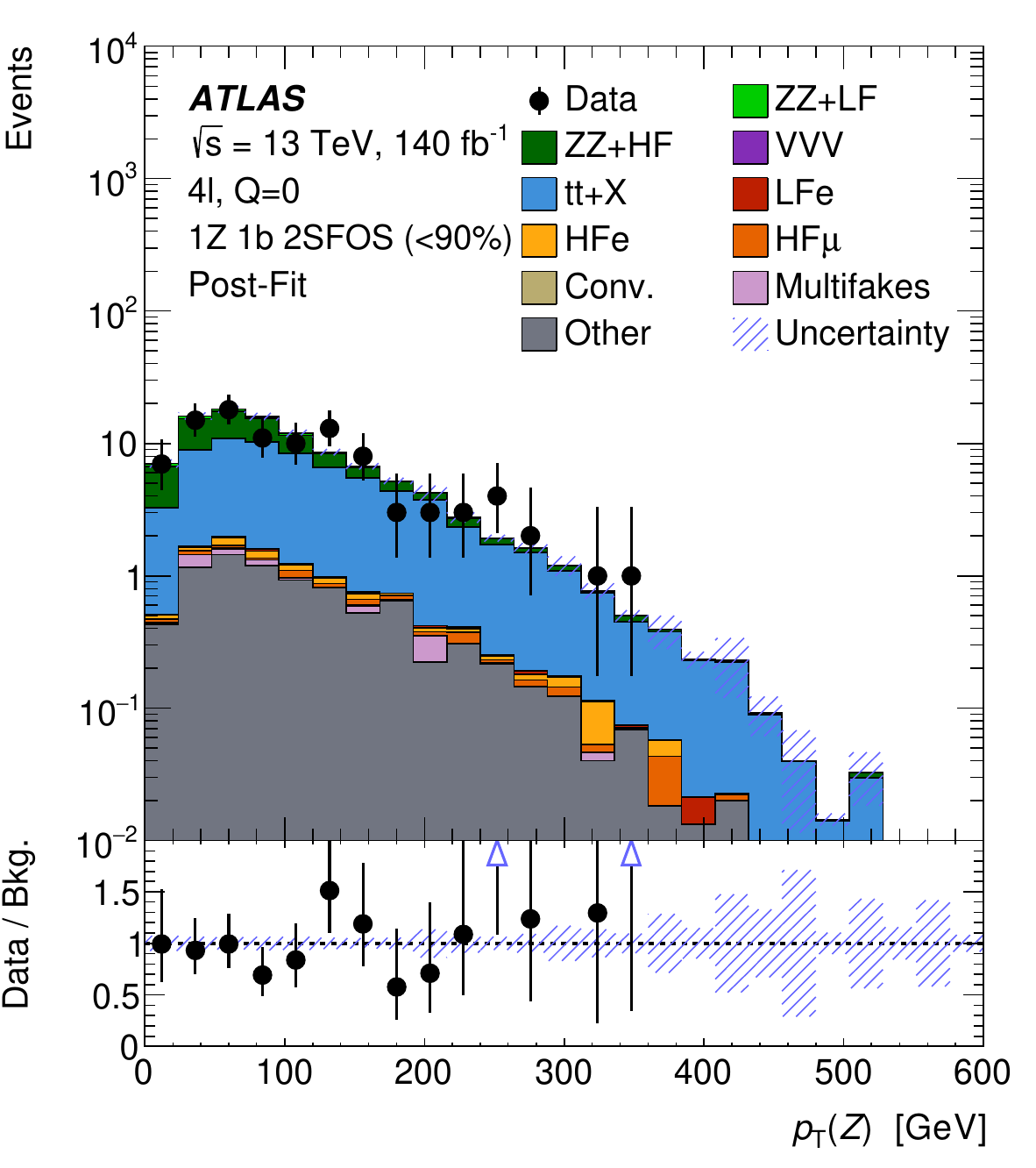}}
\subfloat[]{\includegraphics[width=0.32\textwidth]{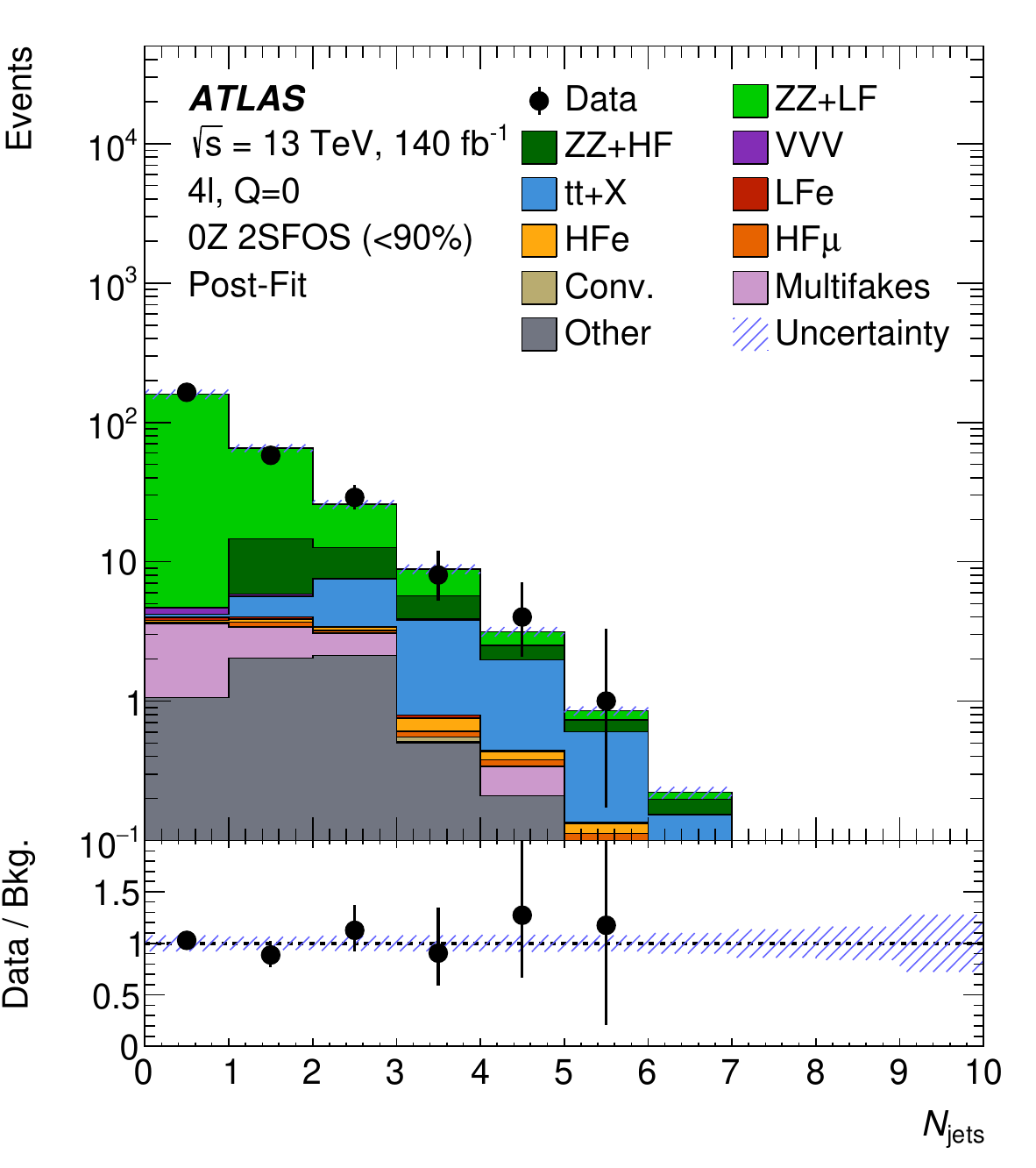}}
\caption{
Comparison between data and the background prediction for the (a) \mtfourlmet, (b) \mthreelh, (c) \mZ, (d) \met, (e) \ptZ, and (f) \njets distribution in the (a, d) 2Z 0b, (b, e) 1Z 1b 2SFOS, and (c, f) 0Z 2SFOS region, after requiring the anomaly score to be below the 90\% background rejection point.
The background contributions after the likelihood fit to data (`post-fit') for the background-only hypothesis are shown as filled histograms.
The `tt+X' background component includes the \ttZ, and \ttH processes.
The `HF $\ell$' (`LF $\ell$') background component refers to processes containing one non-prompt light lepton from heavy-flavour (light-flavour) hadron decays.
The ratio of the data to the background prediction (`Bkg.') is shown in the lower panel.
The `Other' contribution is dominated by the $tWZ$ production.
The size of the combined statistical and systematic uncertainty in the background prediction is indicated by the blue hatched band.
The upward-pointing blue arrows indicate points for which the data-to-background (`Data/Bkg.’) ratio exceeds the vertical range of the figure.
The last bin contains the overflow.
} \label{fig:background_ZZ_ttZ_inputVar}
\end{figure}

A data-driven correction to the \VV jet multiplicity spectrum is derived from a \WZ region enriched in \WZ + LF jets, following the procedure described in Ref.~\cite{EXOT-2021-31}. The WZ/ttZ CR is used in the likelihood fit to improve the prediction of the background contribution from the $\WZ$ + LF jets, $\WZ$ + HF jets, and $\ttdy$ processes. The number of $b$-jets provides good discrimination in the WZ/ttZ CR between the \WZ and \ttdy processes and is the variable used in this region in the fit.
The low-anomaly-score regions, as defined in Section~\ref{sec:anomaly}, are enriched in \ZZ background and provide constraints on the \ZZ background process, where different normalisation factors are assigned to the contributions from \ZZ + LF jets ($\ZZLFzeroZ$, $\ZZLFoneZ$, $\ZZLFtwoZ$), and \ZZ + HF jets.
The splitting of the low-anomaly score discovery regions \llll \qzero 1Z 2SFOS and \llll \qzero 2Z based on the $b$-jet multiplicity allows further discrimination between \ZZ + LF jets against \ttZ and \ZZ + HF jets.
The modelling of the dominant background processes was validated using events with low-anomaly scores. Figure~\ref{fig:background_ZZ_ttZ_inputVar} shows the modelling of some of the input variables to the AD training in regions enriched in \ZZ and \ttdy processes.

Non-prompt leptons originate from material conversions, LF and HF hadron decays, or the improper reconstruction of other particles, and their relative composition depends on the lepton quality requirements and event categories. These backgrounds are estimated from simulation, with the normalisation determined by the likelihood fit.
The main contribution to the non-prompt-lepton background is from $\ttbar$ production, with smaller contributions from $V$+jets, single-top-quark, \ttW, and \WZ processes. %
A correction based on theoretical predictions at NNLO QCD and NLO EW~\cite{Czakon_2017} is applied to the \ttbar process to improve the distributions of the \pT and of the invariant mass of the $t\bar{t}$ system ($m(t\bar{t})$) at parton level.
The non-prompt leptons in the simulated samples are labelled according to whether they originate from HF or LF hadron decays, or from a material conversion candidate. The HF category includes leptons from both bottom and charm decays.
In the discovery/benchmark regions there are additional contributions from backgrounds with more than one non-prompt lepton (referred to as ``multifakes''). The non-prompt lepton composition in multifakes varies depending on the region, and the presence of each non-prompt lepton type is corrected with the corresponding normalisation factor.

Normalisation factors for four non-prompt-lepton background contributions (HF non-prompt electron, HF non-prompt muon, LF non-prompt electron, and electron from internal or material conversion) are estimated from the likelihood fit to data.
The measured values of the six prompt-lepton and four non-prompt-lepton background normalisation factors in the fit to data for the background-only hypothesis using the CRs and the low-anomaly score discovery regions are:

${\lambda}_\textrm{\WZ + LF} = 0.98 \pm 0.09$,
${\lambda}_\textrm{\WZ + HF} = 1.15 \pm 0.31$,
${\lambda}_\textrm{\ZZLFzeroZ} = 0.87 \pm 0.07$,
${\lambda}_\textrm{\ZZLFoneZ} = 1.01 \pm 0.04$,
${\lambda}_\textrm{\ZZLFtwoZ} = 1.04 \pm 0.04$,
${\lambda}_\textrm{\ttZ} = 1.31 \pm 0.11$,
${\lambda}^\textrm{heavy}_{e} = 0.95 \pm 0.06$,
${\lambda}^\textrm{heavy}_{\mu} = 0.95 \pm 0.04$,
${\lambda}^\textrm{light}_{e} = 0.66 \pm 0.09$, and
${\lambda}^\textrm{conv}_e = 1.03 \pm 0.09$.

\begin{figure}[tb]
\centering
\subfloat[]{\includegraphics[width=0.35\textwidth]{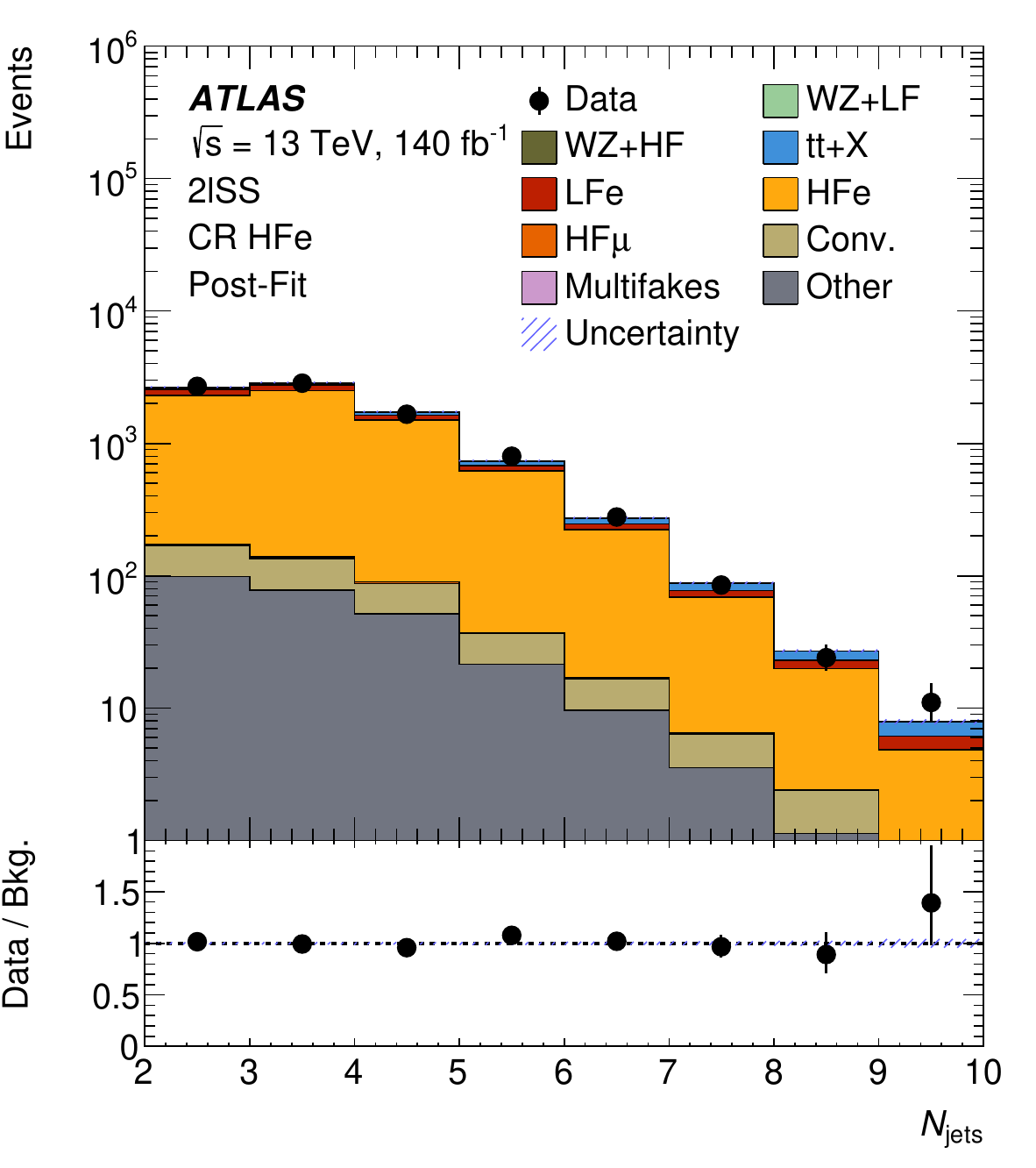}}
\subfloat[]{\includegraphics[width=0.35\textwidth]{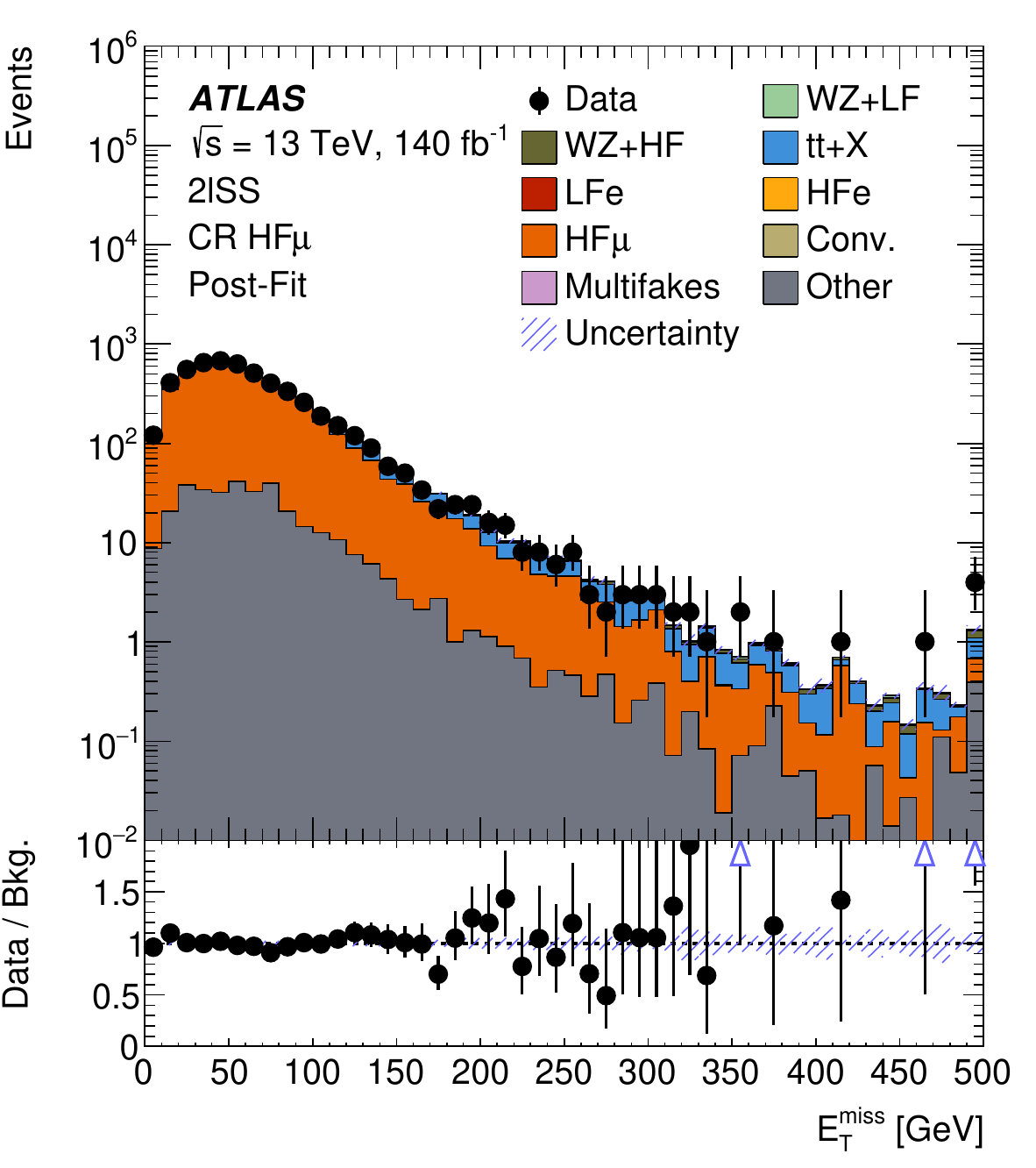}} \\
\subfloat[]{\includegraphics[width=0.35\textwidth]{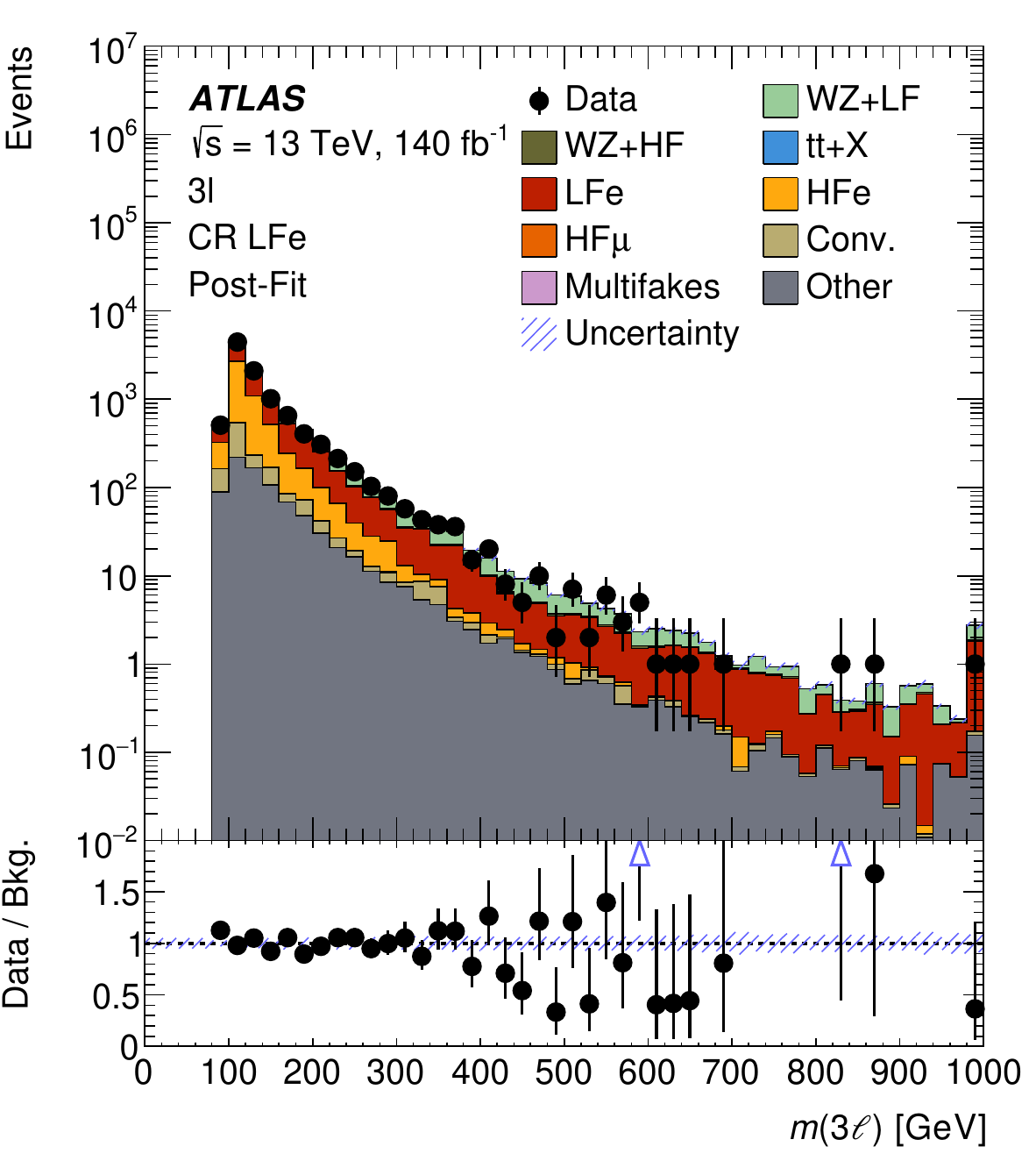}}
\subfloat[]{\includegraphics[width=0.35\textwidth]{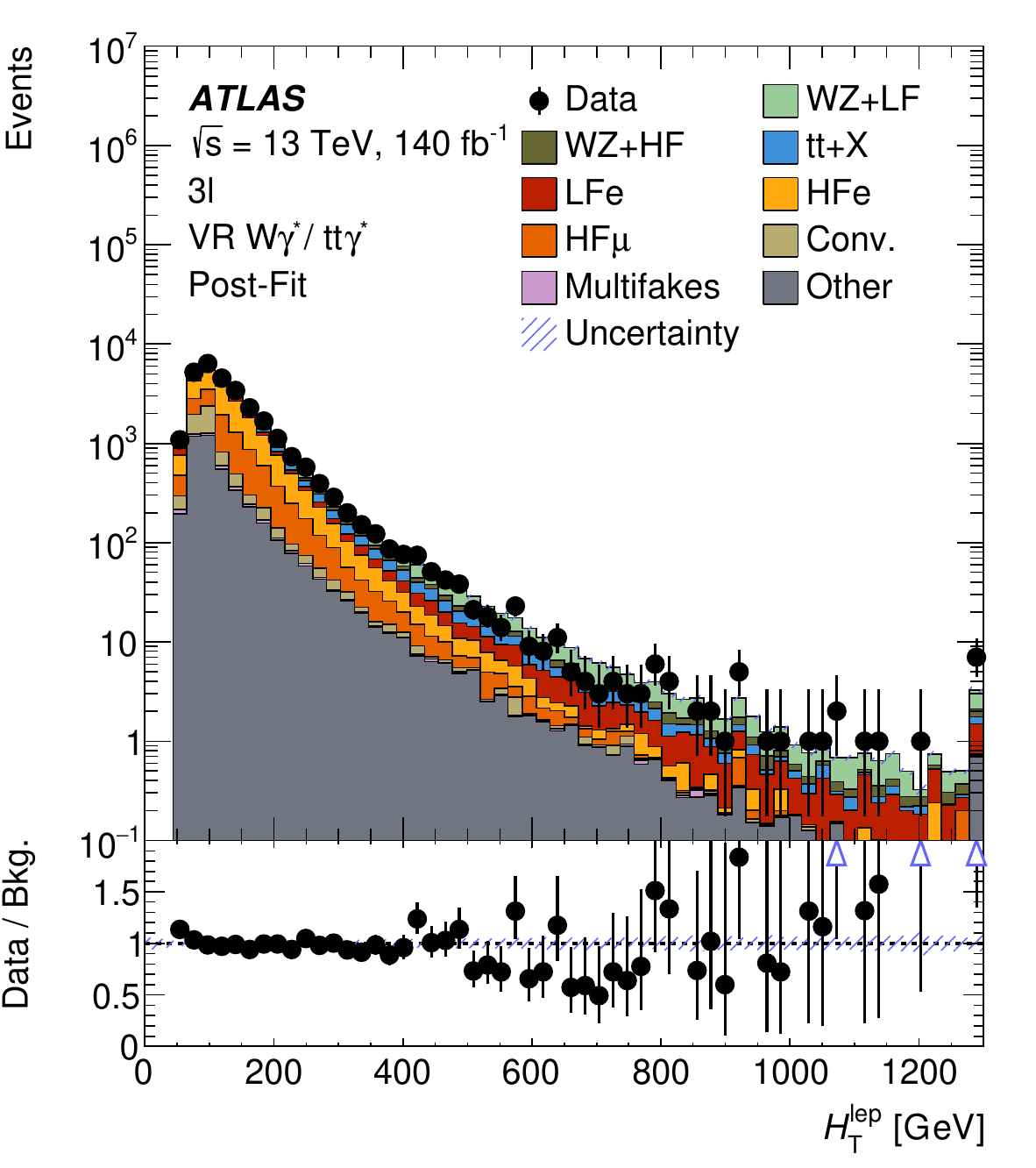}}
\caption{
Comparison between data and the background prediction for the (a) \njets, (b) \met, (c) \mthreel, and (d) \HTlep distribution in the (a) HFe CR, (b) HF$\mu$ CR, (c) LFe CR, and (d) the W$\gamma^*$/$t\bar{t}\gamma^*$ validation region.
The background contributions after the likelihood fit to data (`post-fit') for the background-only hypothesis are shown as filled histograms.
The ratio of the data to the background prediction (`Bkg.') is shown in the lower panel.
The `tt+X' background component includes the \ttW, \ttZ, and \ttH processes.
The `Other' contribution in the $2\ell SS$ regions is dominated by the $\ttbar$ and \ttW production, and in the $3\ell$ regions is dominated by the \ZZ + LF production.
The size of the combined statistical and systematic uncertainty in the background prediction is indicated by the blue hatched band.
The upward-pointing blue arrows indicate points for which the data-to-background (`Data/Bkg.’) ratio exceeds the vertical range of the figure.
The last bin contains the overflow.
} \label{fig:background_fakes_inputVar}
\end{figure}

Kinematic variables were constructed in the fake-enriched CRs in order to validate the modelling of inputs to the AD from fake processes.
Figure~\ref{fig:background_fakes_inputVar} shows representative variables in the HFe CR, HF$\mu$ CR, LFe CR, and the W$\gamma^*$/$t\bar{t}\gamma^*$ validation region, where good background modelling is seen after the fit to data. The latter validation region is defined as the WZ/ttZ CR with the exception that a $Z$ boson candidate veto is applied. It is used to validate the modelling of the electron conversion background.

Backgrounds with leptons with the charge incorrectly assigned affect primarily the $\llSS$ and \llll \qtwo channels and predominantly arise from $t\bar{t}$ production, where one electron undergoes a hard bremsstrahlung and an asymmetric conversion ($e^\pm \to e^\pm \gamma^* \to e^\pm e^+ e^-$) or a mismeasured track curvature.
The muon charge misassignment rate is negligible in the $\pt$ range relevant to this analysis. The electron charge misassignment rate is measured in data using samples of $Z \to e^+e^-$ events reconstructed as same-charge pairs and as opposite-charge pairs, with the background subtracted via a sideband method~\cite{EGAM-2018-01}.
The charge misassignment rate is parameterised as a function of electron $\pt$ and $|\eta|$. It varies from about $10^{-4}$ for low-$\pt$ electrons ($17 \leq \pt \leq 50~\gev$) that satisfy $|\eta| \leq 1.37$, to about $3 \times 10^{-2}$ for high-$\pt$ electrons ($\pt \geq 200~\gev$) in the region $2 \leq  |\eta| \leq 2.47$.
To estimate the electron QMisID background in each of the corresponding event categories, the measured charge misassignment rate is then applied to data events satisfying the requirements of the $\llSS$ and \llll \qtwo channels, except that the total lepton charge is required to be zero. %


%

%
\section{Systematic uncertainties}
%

The search sensitivity is limited by the number of data events rather than by the systematic uncertainties in the background estimate.
The uncertainty in the measurement of the combined 2015--2018 integrated luminosity is 0.83\%~\cite{DAPR-2021-01}, obtained using the LUCID-2 detector~\cite{LUCID2} for the primary measurements, complemented by the ones using the inner detector and calorimeters.

Uncertainties associated with the lepton selection arise from the trigger, reconstruction, identification and isolation efficiencies, and the lepton momentum scale and resolution~\cite{PERF-2015-10,EGAM-2018-01,MUON-2022-01}.

Uncertainties associated with the jet selection arise from the JES, the JVT requirement and the JER~\cite{JETM-2018-05,PERF-2014-03}. The JES and its uncertainties are derived by combining information from test-beam data, collision data and simulation~\cite{JETM-2018-05}. The JES (JER) have 30 (13) components included in the fit. The uncertainties in the JES, JER and JVT increase at lower jet \pt.
The efficiency of the flavour-tagging algorithm is measured for each jet flavour using control samples in data and in simulation. From these measurements, correction factors are derived to correct the tagging rates in the simulation~\cite{FTAG-2018-01,FTAG-2020-08,FTAG-2019-02}. Experimental uncertainties in these correction factors are taken as uncorrelated between $b$-jets, $c$-jets, and
light-flavour jets. An additional uncertainty is assigned to account for the extrapolation of the $b$-tagging efficiency measurement from the \pt region used to determine the correction factors to regions with higher transverse momentum.
Uncertainties on \met are estimated by propagating the uncertainties on the energy and momentum scale of each of the objects contributing to its calculation, and on the soft term resolution and scale~\cite{JETM-2020-03}.

The modelling uncertainties in the main backgrounds are assessed through comparisons with alternative MC samples.
Additional uncertainties are evaluated from renormalisation and factorisation scale variations by a factor of 0.5 and 2, relative to the nominal scales, for the \ttbar, \Zjets, \ttW, \ttZ, and diboson samples. An additional 20\% uncertainty is assigned to the \ttW electroweak contribution~\cite{PhysRevD.103.094014}.

For the \ttbar process, four additional uncertainties are considered related to the reweighting method itself, derived by comparing the nominal reweighted SM \ttbar\ sample to a sample obtained through an alternative reweighting, where the renormalisation and the factorisation scales are separately varied by a factor of 0.5 and 2 and the reweighting is applied on the (anti-)top quark \pT or on the $m(t\bar{t})$ distribution independently.

All alternative \ttbar MC samples are reweighted to the same higher-order predictions as the nominal \POWHEG+\PYTHIA[8.230] MC sample.
In addition to the comparison to the alternative MC sample \POWHEGBOX+\HERWIG[7.1.3], the nominal predictions are also compared with those obtained from an alternative sample generated as the nominal sample but setting the $p_{\mathrm{T}}^{\mathrm{hard}}$ parameter in \PYTHIA to 1 instead of 0~\cite{ATL-PHYS-PUB-2023-029}. This parameter regulates the definition of the vetoed region of the showering to avoid holes or overlaps in the phase space filled by \POWHEG and \PYTHIA.
An uncertainty related to the choice of the \hdamp parameter is estimated by comparing the predictions of the nominal sample to those obtained with an alternative sample with the \hdamp parameter increased by a factor of 1.5 compared with its nominal value. %
Variations in the initial-state radiation (ISR) are estimated by varying the factorisation and renormalisation scales independently up and down by a factor of two.
Similarly, the uncertainty related to final-state radiation (FSR) is assessed by varying the renormalisation scale for final-state PS emissions up and down by a factor of two.
Finally, the uncertainty associated with the A14 tune is derived by varying the A14 tune (Var3c), which affects the renormalisation scale variations in the ISR PS\@. No theory reweighting is applied to this systematic.

For the \Zjets process, the uncertainty related to the upper cut-off of perturbative calculations for PS evolution is known as the re-summation scale. This uncertainty is evaluated at truth level by varying the nominal value of 2 GeV by a factor of 4 up and $1/4$ down. Fiducial cuts are applied at truth level to define a phase space close to that used at reconstruction level in the discovery/benchmark regions where \Zjets is a dominant background. Additionally, the uncertainty related to the choice of the CKKW merging scale, i.e. the scale for calculating the overlap between jets from the ME and the PS, is derived similarly; the nominal value of 20 GeV is varied down to 15 GeV and up to 30 GeV and differences relative to the nominal distribution are evaluated at truth level.

The statistical uncertainty in the fitted parameters for the $VV$ jet-multiplicity correction is propagated as an uncertainty in the diboson background. Finally, additional normalisation uncertainties are included for all processes whose normalisation is not obtained from the fit. In particular, for the \tttt, \ttH, and $tZ$ processes, cross-section uncertainties of 20\%~\cite{Frederix:2017wme}, 11\%~\cite{deFlorian:2016spz}, and 5\%~\cite{TOPQ-2018-01} are assigned, respectively, while for \ttt, $tWZ$, \ttWW, and triboson backgrounds a 50\% cross-section uncertainty is assigned as a conservative estimate, since they are small backgrounds and have low impact on the search.
Additional modelling uncertainties are considered for \ttW, \ttH, and \tttt by comparing the nominal MC event generation described in Section~\ref{sec:datamc} to alternative samples: \Sherpa[2.2.10] (\ttW and \tttt) and \MGNLO and \PYTHIA[8] (\ttH).

Uncertainties in the modelling of the signal samples are evaluated from independently varying the renormalisation and factorisation scale by a factor of 0.5 and 2, relative to the nominal scales.

A systematic uncertainty of 20\% is assigned to the background from electrons with a misidentified charge and accounts for the disagreement observed between the MC simulation for this background and data in the same-sign charge $ee$ inclusive region.


%

%
\section{Results}
\label{sec:results}

A maximum-likelihood fit is performed on all bins in the control regions and discovery/benchmark regions considered in this search as shown in Figure~\ref{fig:sketch_strategy} to simultaneously determine the signal and background yields that are most consistent with the data. In the case of the model-independent fit, five cut-based control regions and ten low-anomaly score discovery regions are fitted together with one signal bin at a time (16 discovery signal bins corresponding to 16 different fits).
The total event yields are used in all regions, except for the WZ/ttZ CR where the \nbjets distribution is used. In the case of the model-dependent fit, the same five cut-based control regions are fitted together with 32 benchmark regions. In each benchmark region, the anomaly score distribution is fitted to data with the exception of the \lllll 0Z benchmark regions, where the total event yield is fitted.

The likelihood function ${\cal L}(\mu,\vec{\lambda}, \vec{\theta})$ is constructed as a product of Poisson probability terms over all bins considered in the search, and depends on: the signal-strength parameter, $\mu$, a multiplicative factor applied to the predicted yield for the probed signal (model-dependent, as detailed in Section~\ref{subsec:modeldepres}) or a dummy signal (model-independent, as detailed in Section~\ref{subsec:modelindepres}); $\vec{\lambda}$, the normalisation factors for several backgrounds; $\vec{\theta}$, a set of nuisance parameters (NPs), encoding systematic uncertainties in the signal and background expectations~\cite{Cranmer:2012sba}.
Both $\mu$ and $\vec{\lambda}$ are treated as free parameters in the likelihood fit. The NPs $\vec{\theta}$ allow variations of the expectations for signal and background according to the systematic uncertainties, subject to Gaussian or Poisson constraints in the likelihood fit.
Statistical uncertainties in each bin due to the limited size of the simulated samples are taken into account by dedicated parameters using the Beeston--Barlow `lite' technique~\cite{Conway:2011}.

The test statistic $q_\mu$ is defined as the profile likelihood ratio:
$q_\mu = -2\ln({\cal L}(\mu,{\hat{\vec{\lambda}}}_\mu, {\hat{\vec{\theta}}}_\mu)/{\cal L}(\hat{\mu},\hat{\vec{\lambda}}_{\hat{\mu}}, \hat{\vec{\theta}}_{\hat{\mu}}))$,
where $\hat{\mu}$, $\hat{\vec{\lambda}}_{\hat{\mu}}$, and $\hat{\vec{\theta}}_{\hat{\mu}}$ are the values of the parameters that maximise the likelihood function,
and ${\hat{\vec{\lambda}}}_\mu$ and ${\hat{\vec{\theta}}}_\mu$ are the values of the parameters that maximise the likelihood function for a given value of $\mu$.
The test statistic $q_\mu$ is evaluated with the {\textsc RooFit} package~\cite{RooFit}.
A related statistic is used to determine the probability that the observed data are incompatible with the background-only hypothesis by setting $\mu=0$ in the profile likelihood ratio ($q_0$).
The $p$-value (referred to as $p_0$) representing the probability of the data being compatible with the background-only hypothesis is estimated by integrating the distribution of $q_0$ from background-only pseudo-experiments above the observed value of $q_0$.
Upper limits on the signal production cross-section, calculated for either a dummy signal in the model-independent scenario or for each of the signal scenarios considered in the model-dependent scenario, are derived by using $q_\mu$ in the CL$_{\textrm{s}}$ method~\cite{Junk:1999kv,Read:2002hq}.
For a given signal scenario, values of the production cross-section (parameterised by $\mu$) yielding CL$_{\textrm{s}} < 0.05$ are excluded at $\geq 95\%$ CL, where CL$_{\textrm{s}}$ is computed using pseudo-experiments for the model-independent scenario and the flavourful VLL model-dependent search, and the asymptotic approximation~\cite{Cowan:2010js} for the other model-dependent signal benchmarks.

\subsection{Model-independent results}
\label{subsec:modelindepres}

\begin{figure}[!h]
\centering
\includegraphics[width=\textwidth]{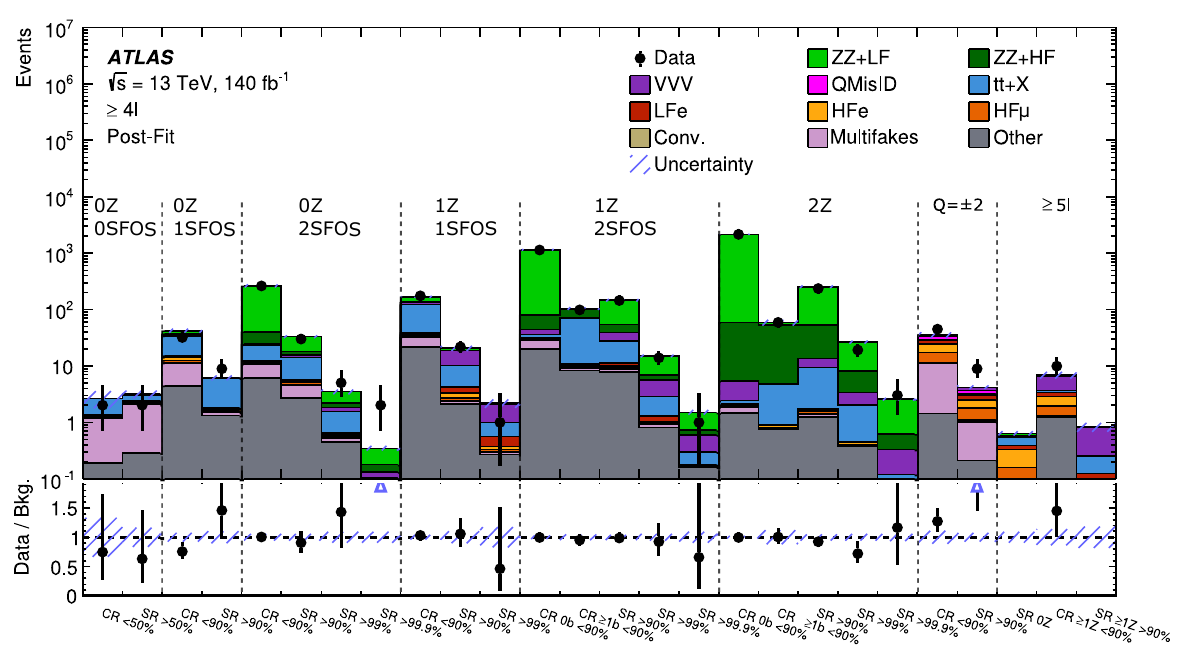}
\caption{
Comparison between data and the background prediction for the discovery regions included in the model-independent fit, after a background-only fit to data in the five cut-based control regions and ten low-anomaly score discovery regions (`post-fit'). All signal bins are shown (`SR') together with the low-anomaly score discovery regions (`CR'). The various requirements on the anomaly score corresponding to the specified background rejection cuts (99.9/99/90/50\%) are shown for all SRs and CRs, except for the \lllll 0Z SR, where no AD is applied.
The lower panels show the ratio of data to the background estimate (`Bkg').
The `tt+X' background component includes the \ttZ (\llll only) and \ttH processes.
The `Other' contribution is dominated by the $VH$ process.
The size of the combined statistical and systematic uncertainty in the background prediction is indicated by the blue hatched band.
The upward-pointing blue arrows indicate points for which the data-to-background (`Data/Bkg.’) ratio exceeds the vertical range of the figure.
\label{fig:Indep_Summary_Scores}}
\end{figure}

Figure~\ref{fig:Indep_Summary_Scores} shows the data and background prediction for the adjacent anomaly score distributions including all signal bins used. No significant deviations from the SM expectations are observed in any of the signal bins considered, with the largest excess seen in the \qtwo and $0Z~2$SFOS $(>99.9\%)$ regions corresponding to 2$\sigma$ local significance each. The global significance is 0.90$\sigma$ and takes into account the look-elsewhere effect. The global significance is estimated through the simulation of a large number of pseudo-experiments according to the background-only expected number of events. Pseudo-experiments are generated using all systematic and statistical uncertainties, by suitably fluctuating each discovery region in a simultaneous and correlated way. Overlaps between discovery regions are further accounted for in the calculation of the look-elsewhere effect.

To calculate an excluded visible cross section, a dummy signal is injected into each signal bin, with a free-floating normalisation parameter (signal strength), denoted $\mu$. The signal bin is then used along with the control regions and low-anomaly score discovery regions in a profile-likelihood fit, and a 95\% confidence limit is found. The excluded visible cross section, $\sigma_{vis}$, is then given by:

\begin{equation}
\sigma_{vis} = \mu \frac{N_{sig}}{\mathcal{L}}
\end{equation}

where $N_{sig}$ is the number of injected dummy signal events and $\mathcal{L}$ is the integrated luminosity.

Figure~\ref{fig:Limits,ModelIndep,FullSysts} shows the expected excluded visible cross section, along with the observed excluded cross section for each model-independent signal bin. The expected significances in each of the model-independent fits to some benchmark signal models (single and doublet \vlle and \vllmu, flavourful \vlle and \vllmu, wino-like chargino and neutralino, and smuon) can be seen in Figure~\ref{fig:Significances,ModelIndep,FullSysts}. The discovery regions with the largest sensitivity to the benchmark models are \llll \qzero 0Z 2SFOS $(>99.9\%)$ and \lllll 0Z for flavourful \vlle and wino-like chargino and neutralino, \llll \qzero $(>99\%)$ for \vllmud, and \lllll 0Z for smuon.

The search is dominated by statistical uncertainties due to the requirement of a large lepton multiplicity: the expected limits obtained with full systematic uncertainties worsen at most 12\% relative to the limits including only statistical uncertainties.

\begin{figure}[!h]
\centering
\includegraphics[width=\textwidth]{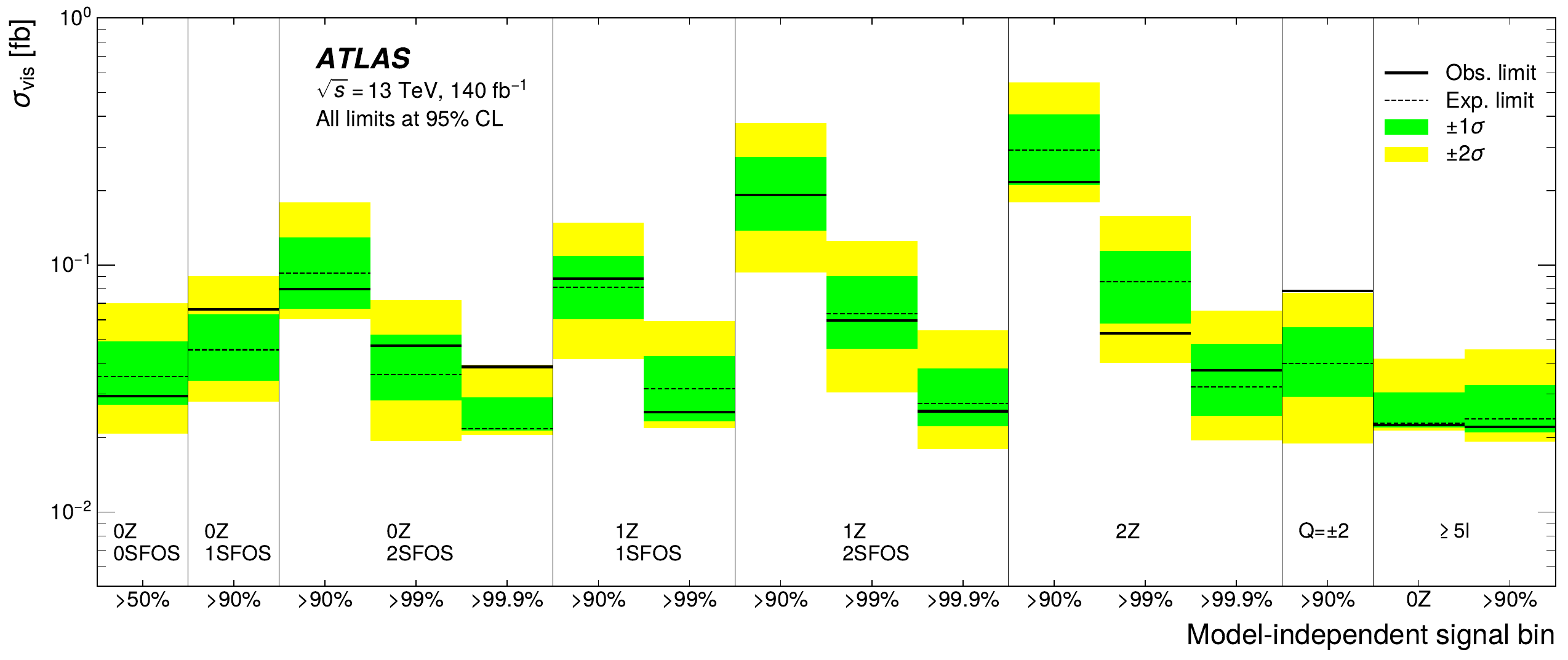}
\caption{Observed and expected excluded visible cross section limits, $\sigma_{\text{vis}}$ for the model-independent regions.
The various requirements on the anomaly score corresponding to the specified background rejection cuts (99.9/99/90/50\%) are shown for all regions, except for the \lllll 0Z SR, where no AD is applied.
\label{fig:Limits,ModelIndep,FullSysts}}
\end{figure}

\begin{figure}[!h]
\centering
\includegraphics[width=\textwidth]{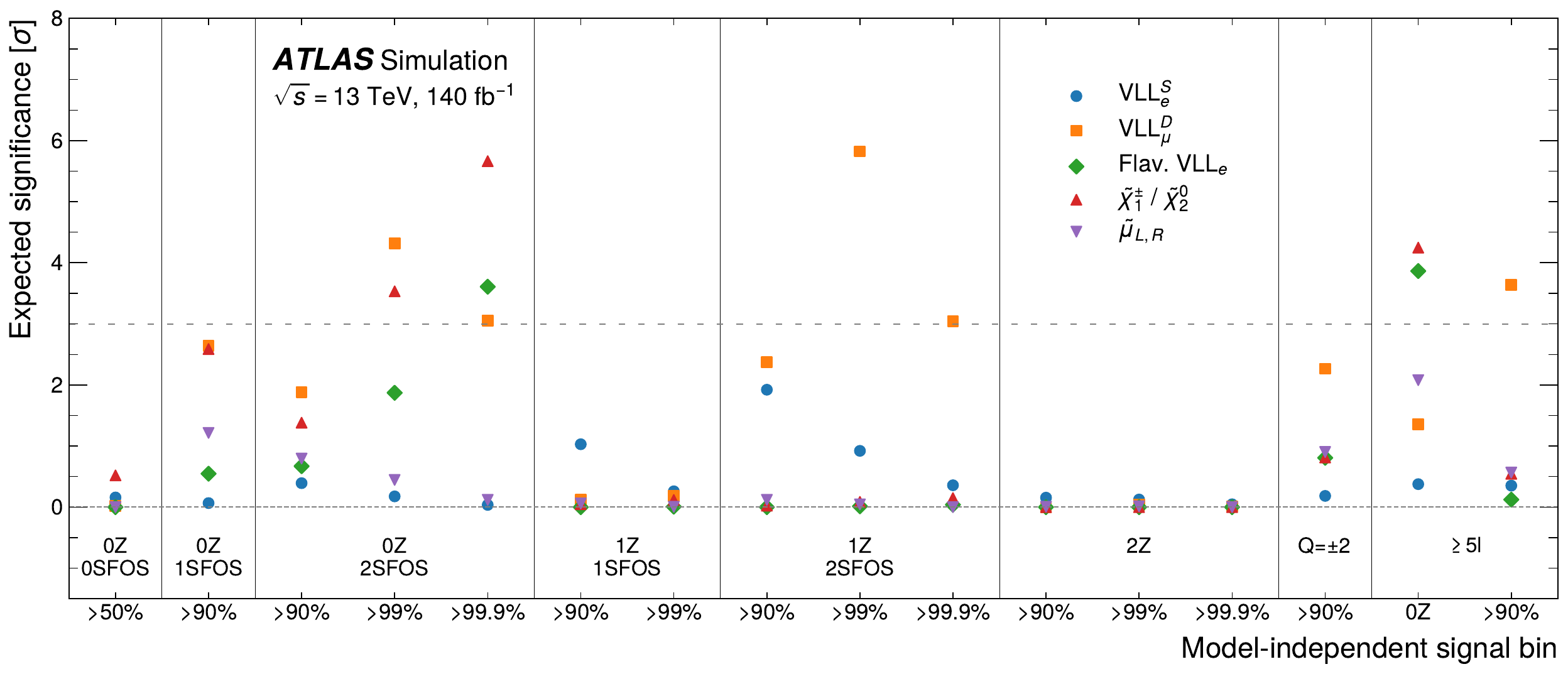}
\caption{Expected significance in the model-independent regions to the benchmark signal models: \vlles with a mass of 200 GeV, \vllmud with a mass of 600 GeV, flavourful \vlle with a mass of 1200 GeV and a scalar $S$ mass of 550 GeV, wino-like chargino with a mass of 1300 GeV and a neutralino mass of 800 GeV, and smuon with a mass of 350 GeV and a neutralino mass of 250 GeV.
The various requirements on the anomaly score corresponding to the specified background rejection cuts (99.9/99/90/50\%) are shown for all regions, except for the \lllll 0Z SR, where no AD is applied.
\label{fig:Significances,ModelIndep,FullSysts}}
\end{figure}

\subsection{Model-dependent results}
\label{subsec:modeldepres}

\begin{figure}[!h]
\centering
\subfloat[]{\includegraphics[width=0.32\textwidth]{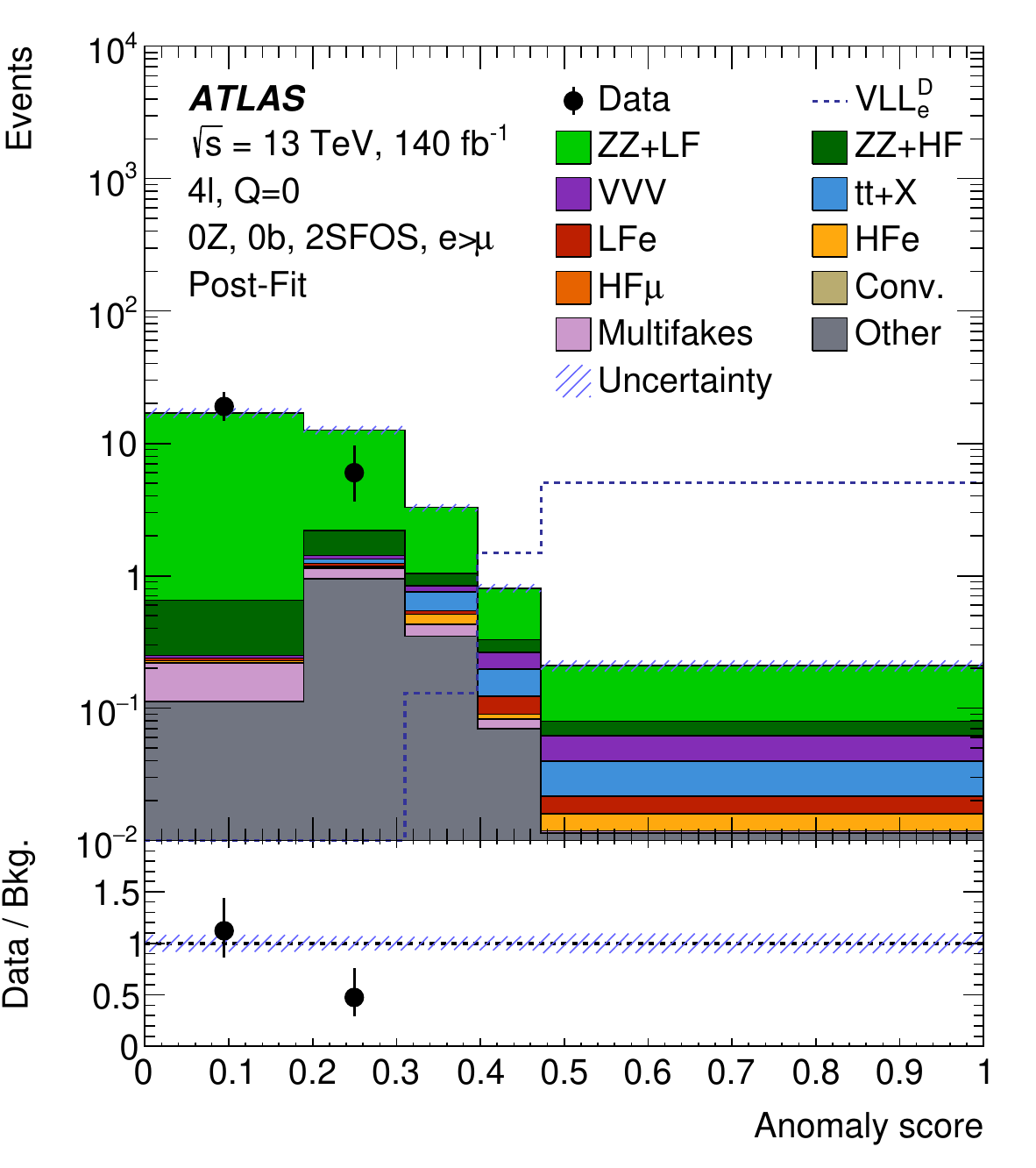}}
\subfloat[]{\includegraphics[width=0.32\textwidth]{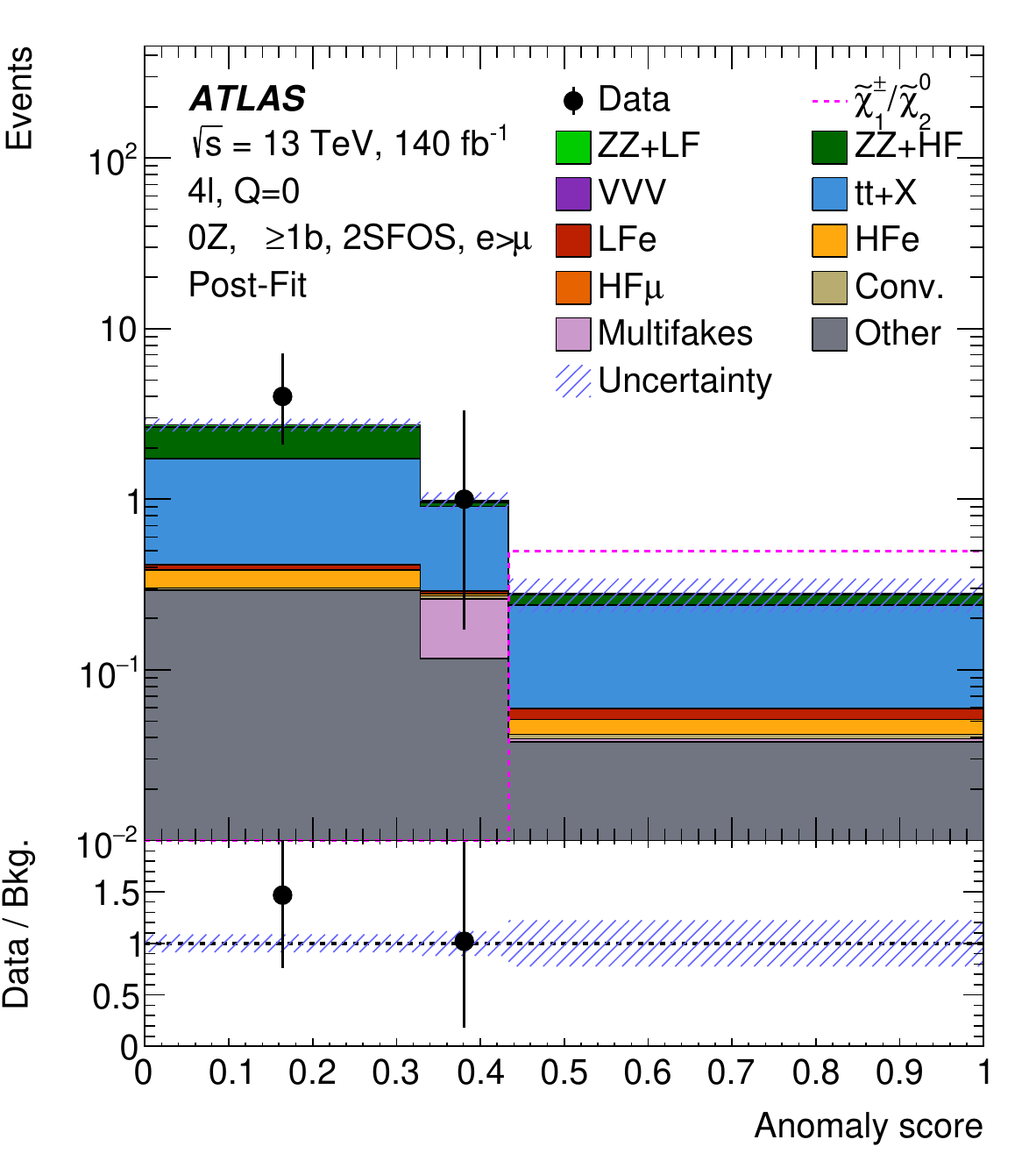}}
\subfloat[]{\includegraphics[width=0.32\textwidth]{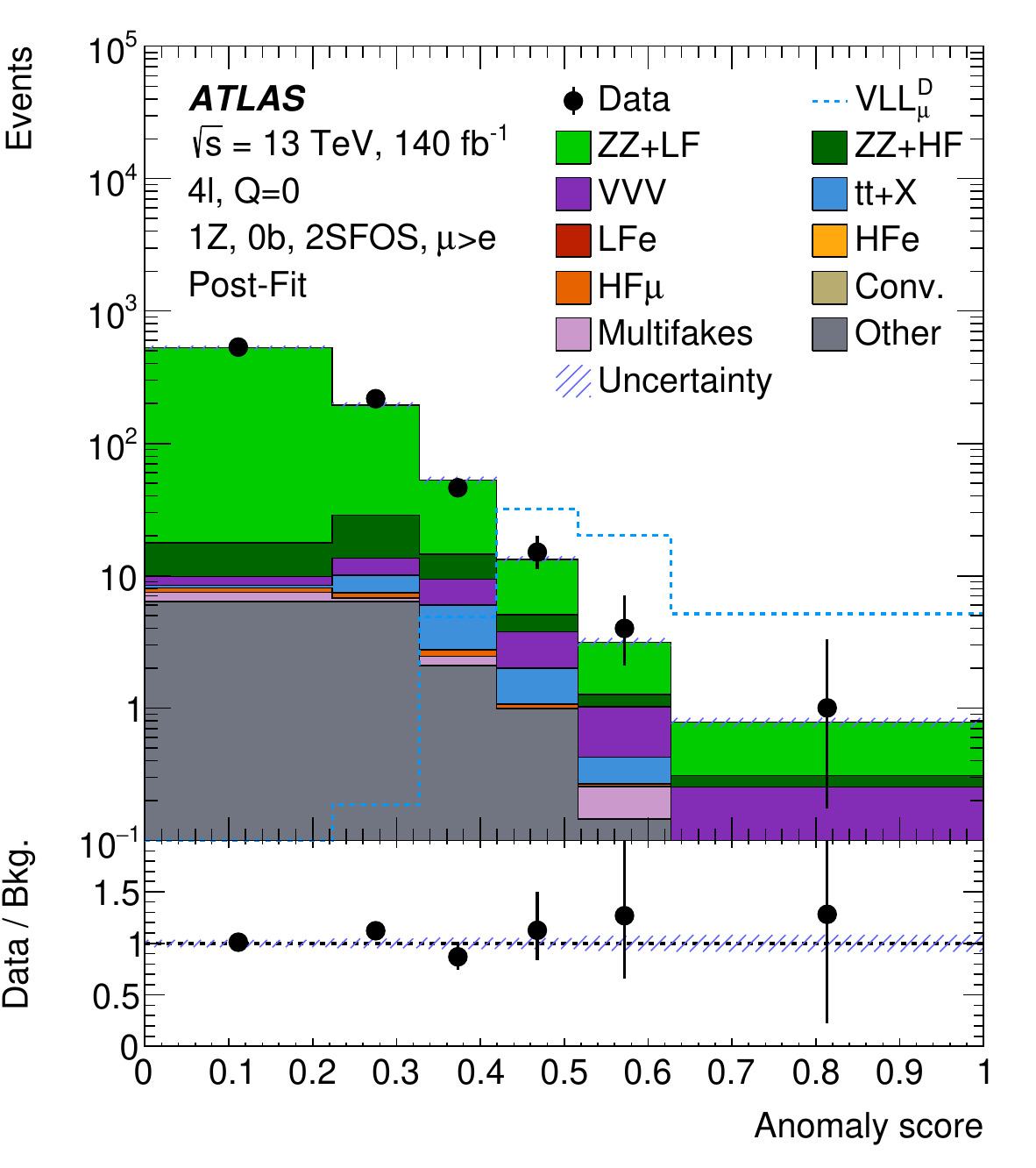}} \\
\subfloat[]{\includegraphics[width=0.32\textwidth]{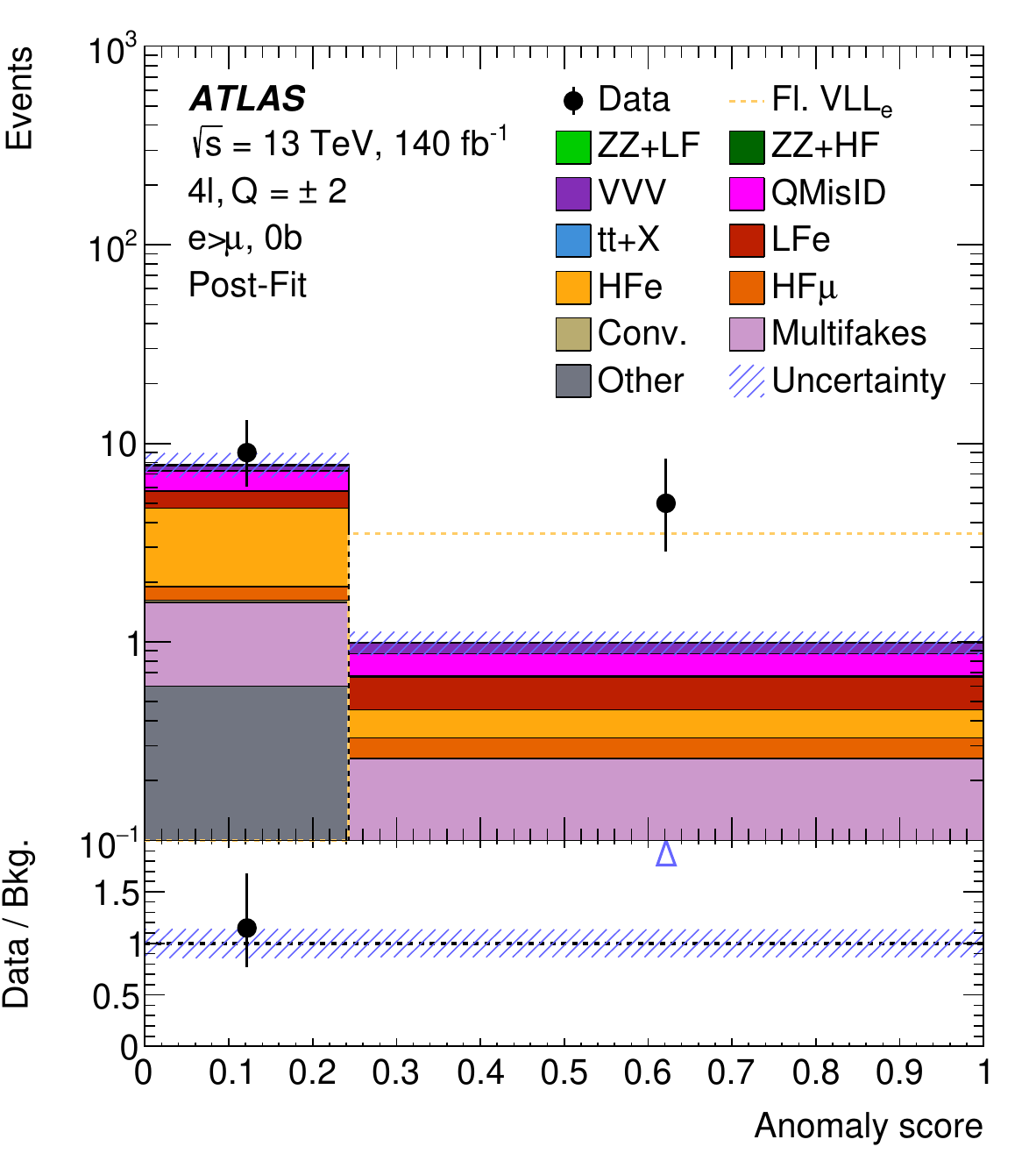}}
\subfloat[]{\includegraphics[width=0.32\textwidth]{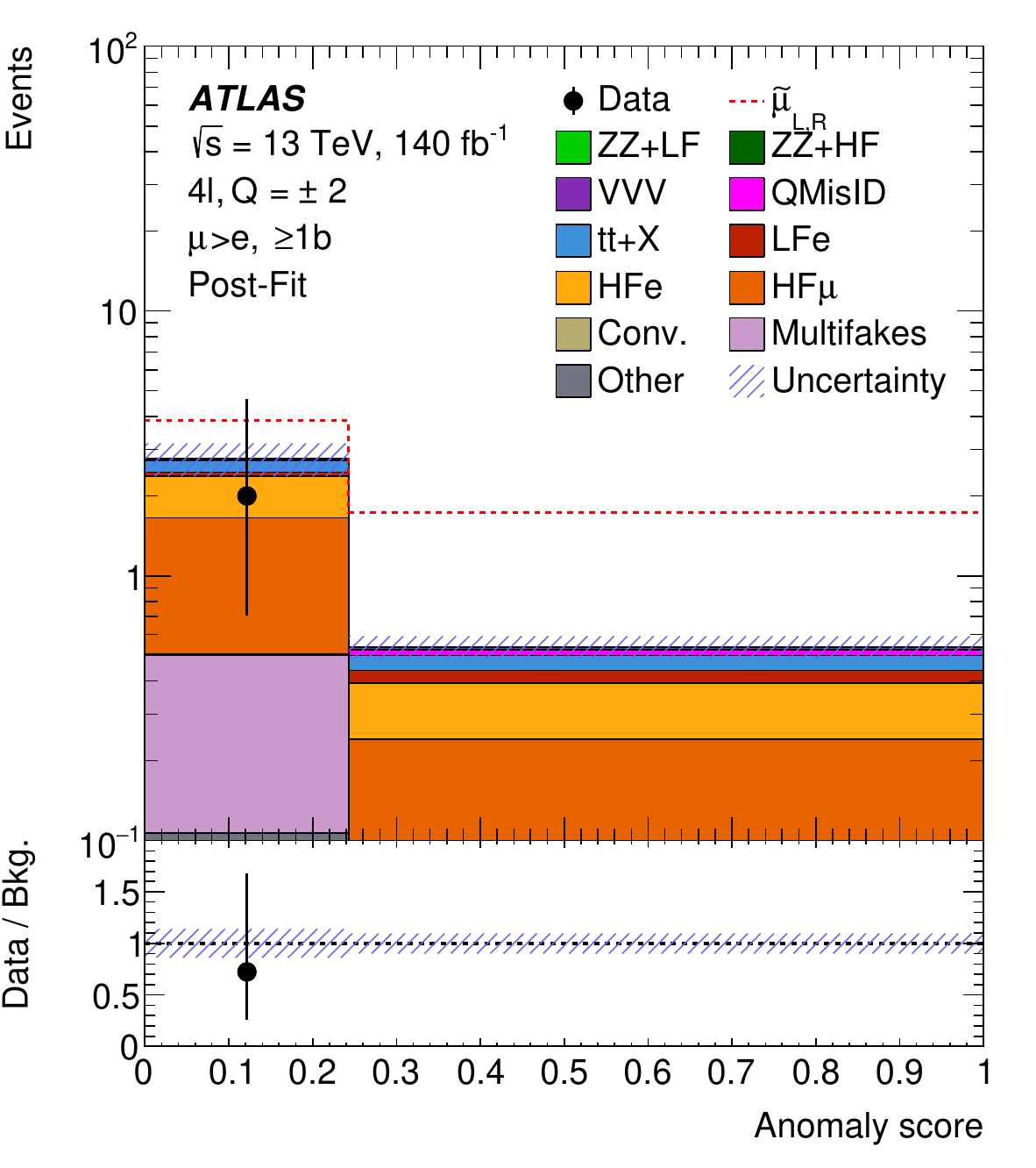}}
\subfloat[]{\includegraphics[width=0.32\textwidth]{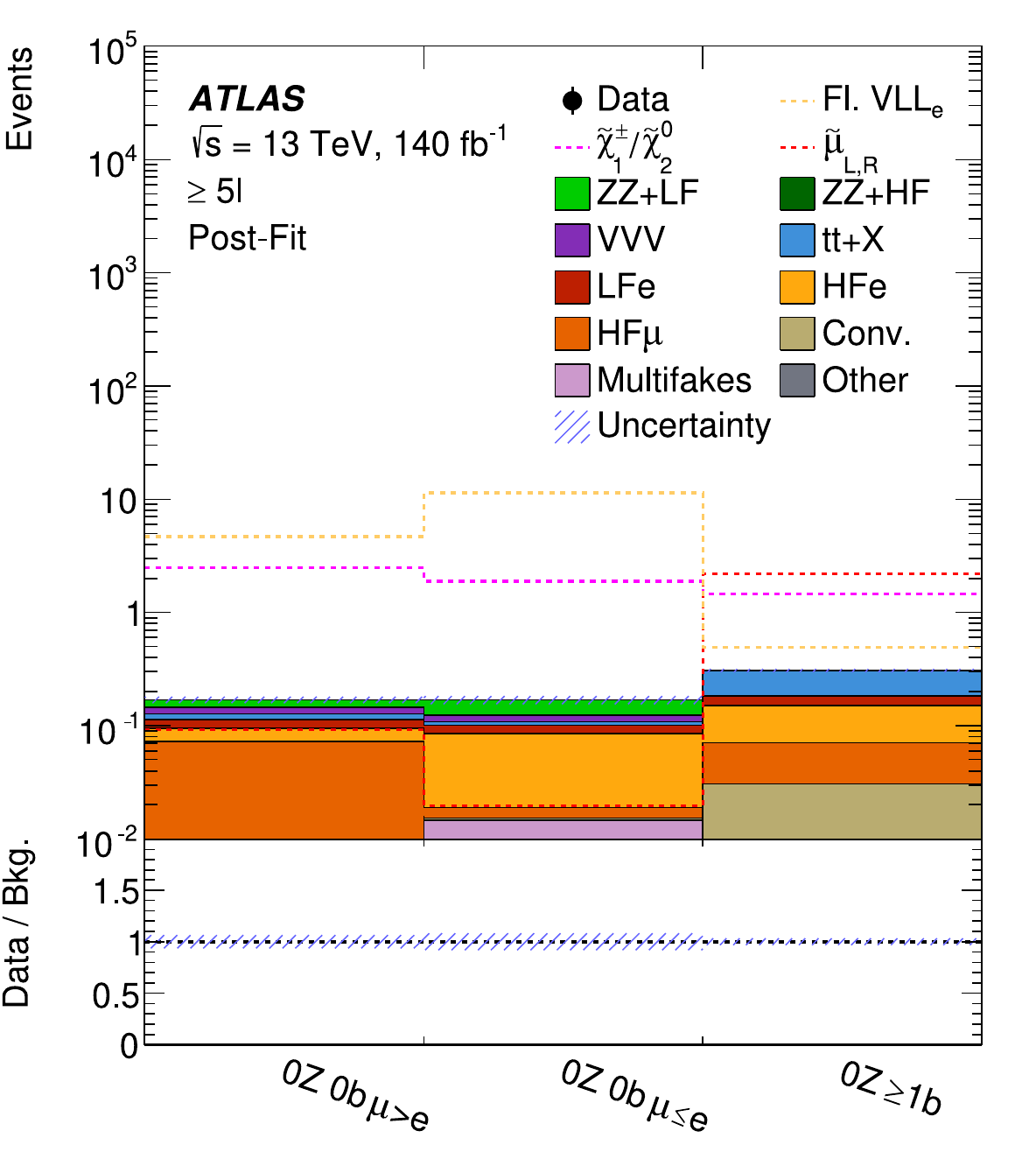}}
\caption{
Comparison between data and the background estimate for the anomaly score distribution used in different benchmark regions, after a background-only fit to data (`post-fit'):
(a) \llll \qzero 0Z 0b 2SFOS $e>\mu$,
(b) \llll \qzero 0Z $\geq$1b 2SFOS $e>\mu$,
(c) \llll \qzero 1Z 0b 2SFOS $\mu>e$,
(d) \llll \qtwo 0b  $e>\mu$,
(e) \llll \qtwo $\geq$1b $\mu>e$, and
(f) the three \lllll regions: 0Z 0b $\mu>e$, 0Z 0b $e\geq \mu$, and 0Z $\geq$1b, defined without AD.
Distributions for a \vlle and \vllmu signal point of mass 500 GeV, one flavourful \vlle signal point for a VLL mass of 1 TeV and scalar of 550 GeV, a wino-like chargino and neutralino signal point of mass 1.3 TeV, and a smuon signal point of mass 350 GeV and neutralino mass of 250 GeV are overlaid for comparison. %
The lower panels show the ratio of data to the background estimate (`Bkg.').
The `tt+X' background component includes the \ttZ (\llll only) and \ttH processes.
The `Other' contribution is dominated by the $VH$ process.
The size of the combined statistical and systematic uncertainty in the signal-plus-background prediction is indicated by the blue hatched band.
The upward-pointing blue arrow indicates a point for which the data-to-background (`Data/Bkg.’) ratio exceeds the vertical range of the figure.
\label{fig:ModelDep_SignalRegions}}
\end{figure}

Comparisons between data and the background prediction for the anomaly score distributions used in a selection of benchmark regions are shown in Figure~\ref{fig:ModelDep_SignalRegions}. Representative benchmark signal models are overlaid in the regions where they are expected to be dominant.
No significant deviations from the SM expectations are observed in any of the benchmark regions considered.
The smallest p-value for the newly probed flavourful VLL is 0.26 for a \vllmu of mass 800~\gev and a scalar mass of 350~\gev, which corresponds to a local significance of 0.63~$\sigma$.

Limits at 95\% CL on the cross-section of relevant benchmark signal models as a function of particle masses are set.
Figures~\ref{fig:Limits,ModelDep,VLLeD}-~\ref{fig:Limits,ModelDep,VLLmuS} show the limits on the cross-section for the models: \vlled, \vllmud, \vlles, and \vllmus. Limits from the dedicated ATLAS search~\cite{EXOT-2021-31} are overlaid, considering only the \llll channel.
The \vlles is observed (expected) to be excluded at $95\%$ CL for masses up to 290 (290) GeV while the \vlled is observed (expected) to be excluded at $95\%$ CL for masses up to 925 (880) GeV. The \vllmus is observed (expected) to be excluded at $95\%$ CL for masses up to 320 (320) GeV while the \vllmud is observed (expected) to be excluded at $95\%$ CL for masses up to 850 (925) GeV.
These limits are more stringent than those from the dedicated analysis considering only the \llll channel where a simple counting analysis is performed.

Figures~\ref{fig:Limits,ModelDep,Fl_VLLe} and~\ref{fig:Limits,ModelDep,Fl_VLLmu} show the first ever limits on the flavourful \vlle or \vllmu mass and the scalar $S$.
The flavourful \vlle is observed (expected) to be excluded at $95\%$ CL for masses up to 1300 (1340) GeV while the flavourful \vllmu is observed (expected) to be excluded at $95\%$ CL for masses up to 1280 (1375) GeV, for all allowed scalar $S$ masses except for the lowest and highest values, where the exclusion can be up to about 50 GeV weaker.

Figure~\ref{fig:Limits,ModelDep,WinoSmuon} shows the limits on the supersymmetric models considered, wino-like chargino and neutralino, and smuon. Wino-like charginos are observed (expected) to be excluded at $95\%$ CL for masses up to 1600 (1600) GeV. Smuon masses up to 495 (530) GeV are observed (expected) to be excluded at $95\%$ CL.
Whereas the limits from the dedicated wino-like chargino and neutralino analysis are more stringent than the current analysis, for the smuon model this result is more sensitive for $\Delta m (\tilde{\mu}_{\mathrm{L,R}}, \tilde{\chi}^0_1)$ larger than 140 GeV compared to the existing expected limits.

\begin{figure}[!h]
\centering
\subfloat[]{\includegraphics[width=0.45\textwidth]{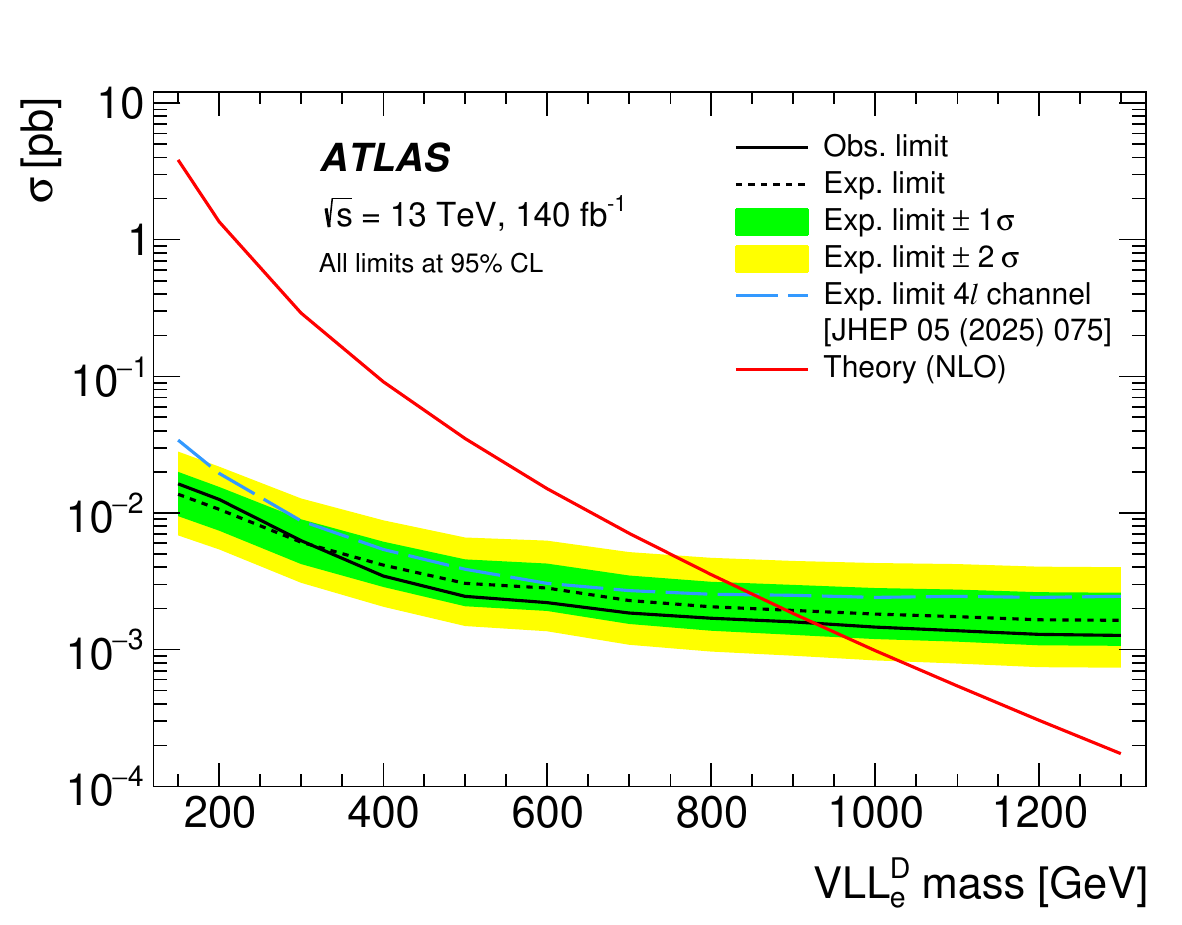}\label{fig:Limits,ModelDep,VLLeD}}
\subfloat[]{\includegraphics[width=0.45\textwidth]{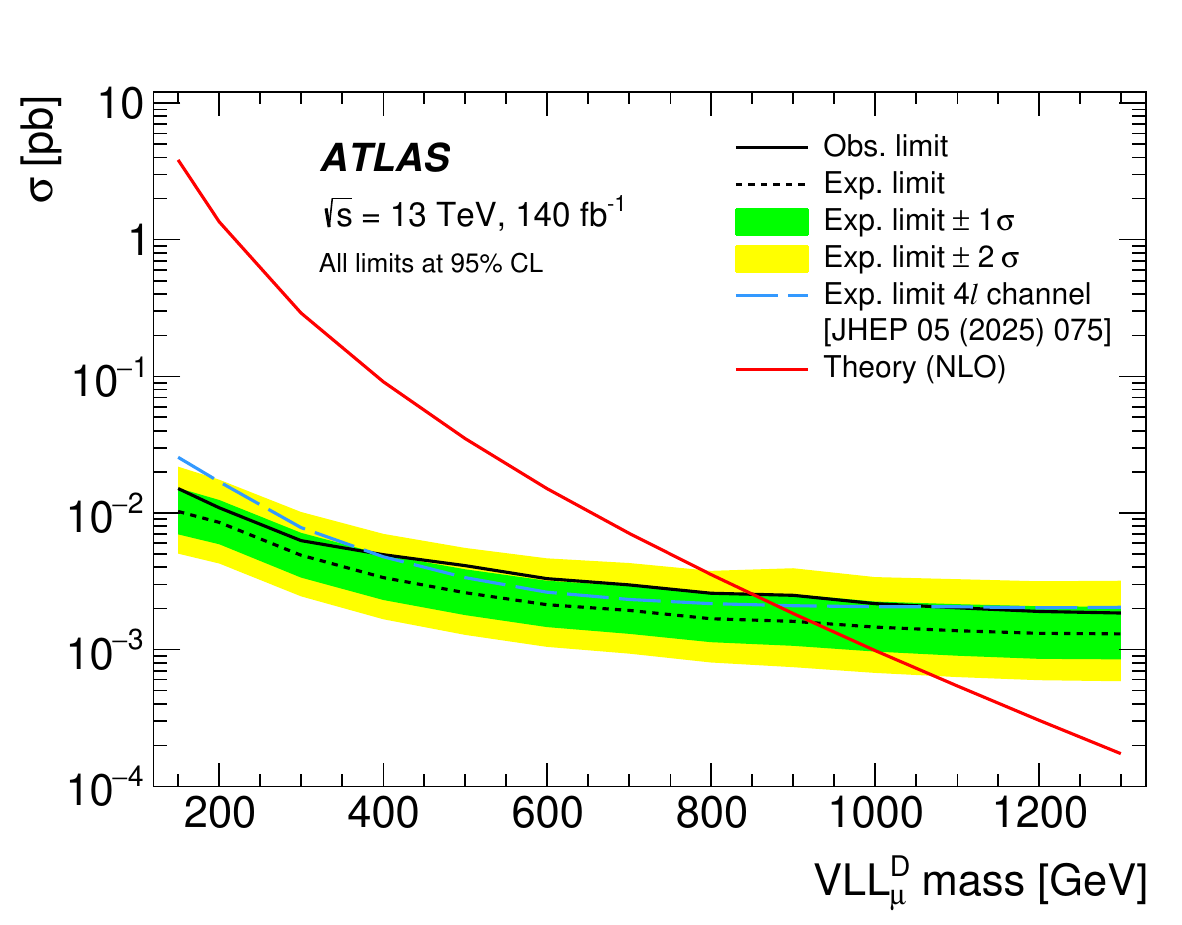}}\\
\subfloat[]{\includegraphics[width=0.45\textwidth]{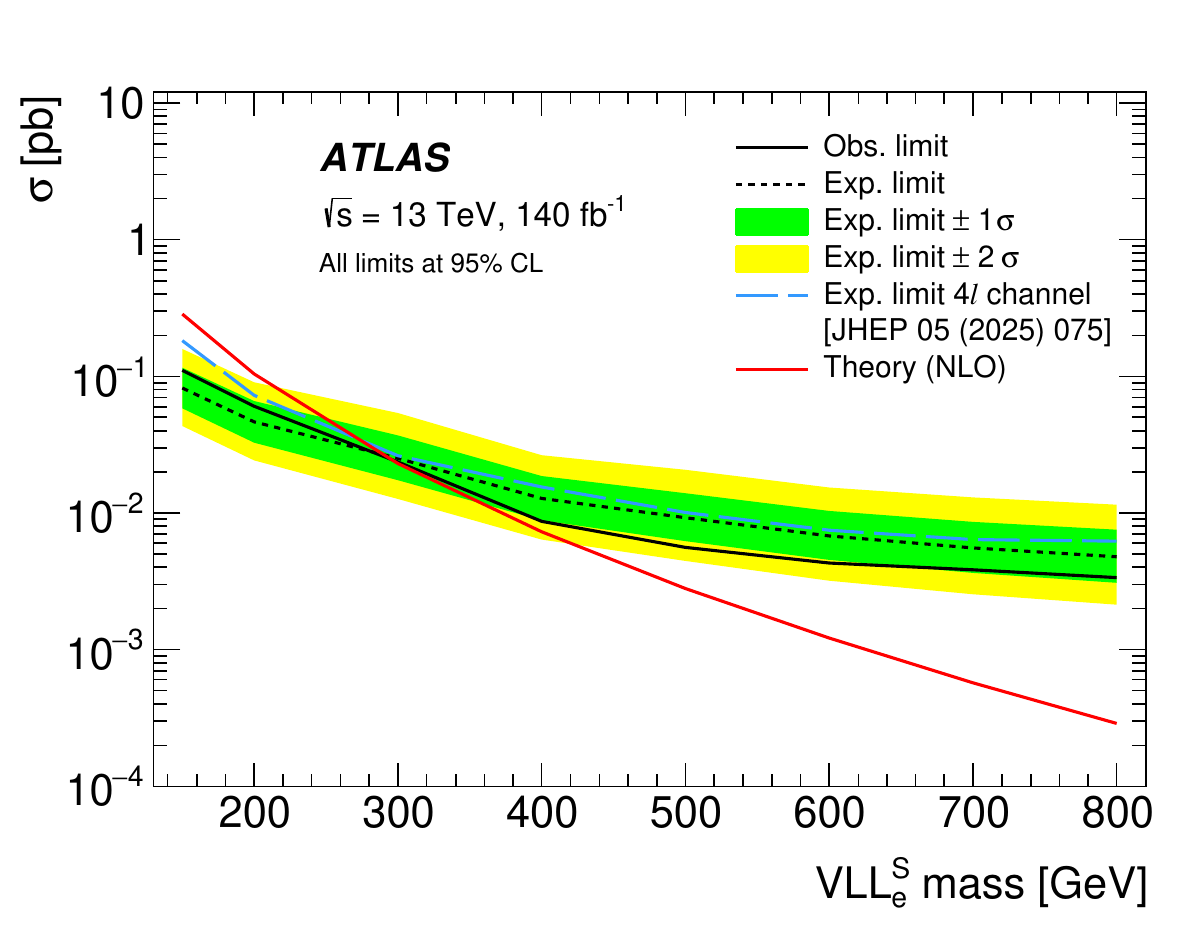}}
\subfloat[]{\includegraphics[width=0.45\textwidth]{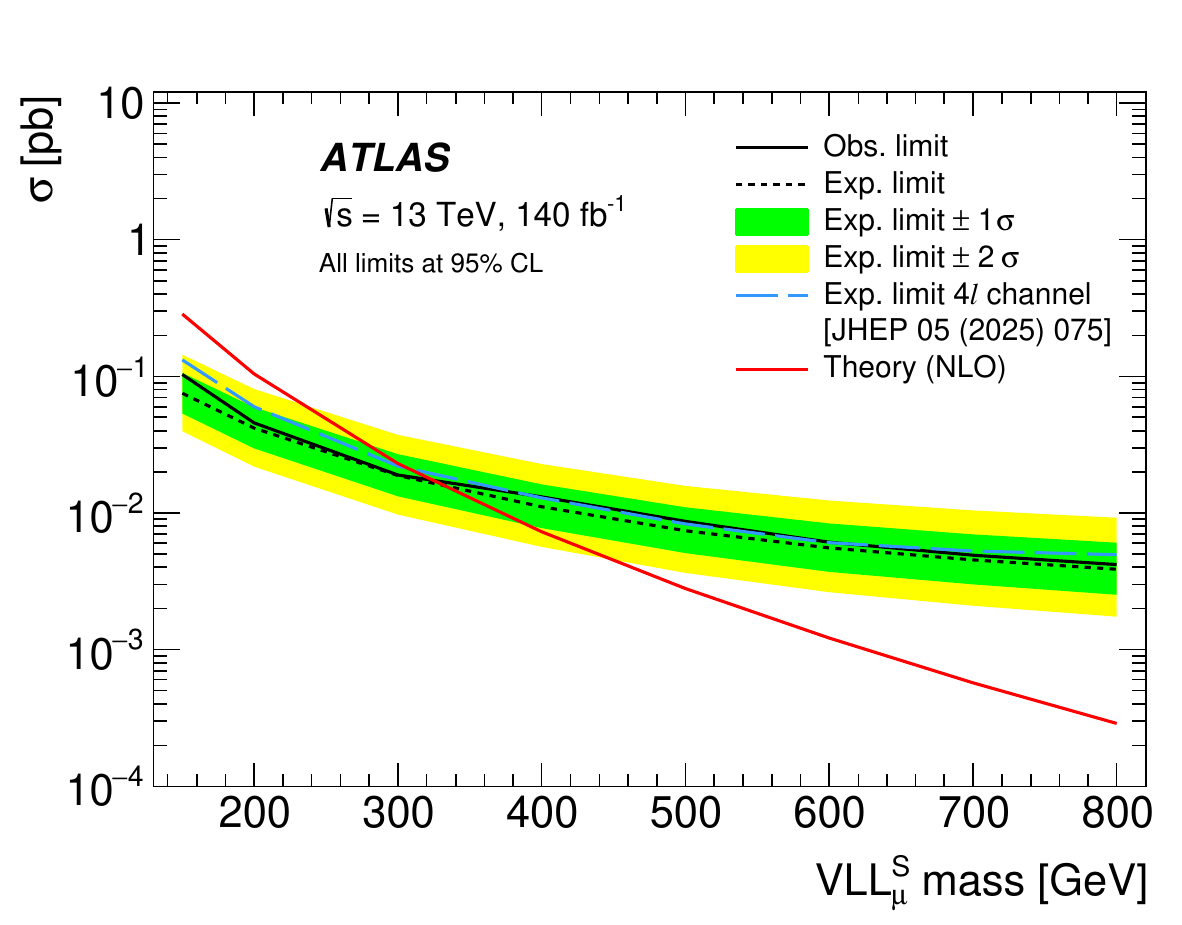}\label{fig:Limits,ModelDep,VLLmuS}}\\
\subfloat[]{\includegraphics[width=0.45\textwidth]{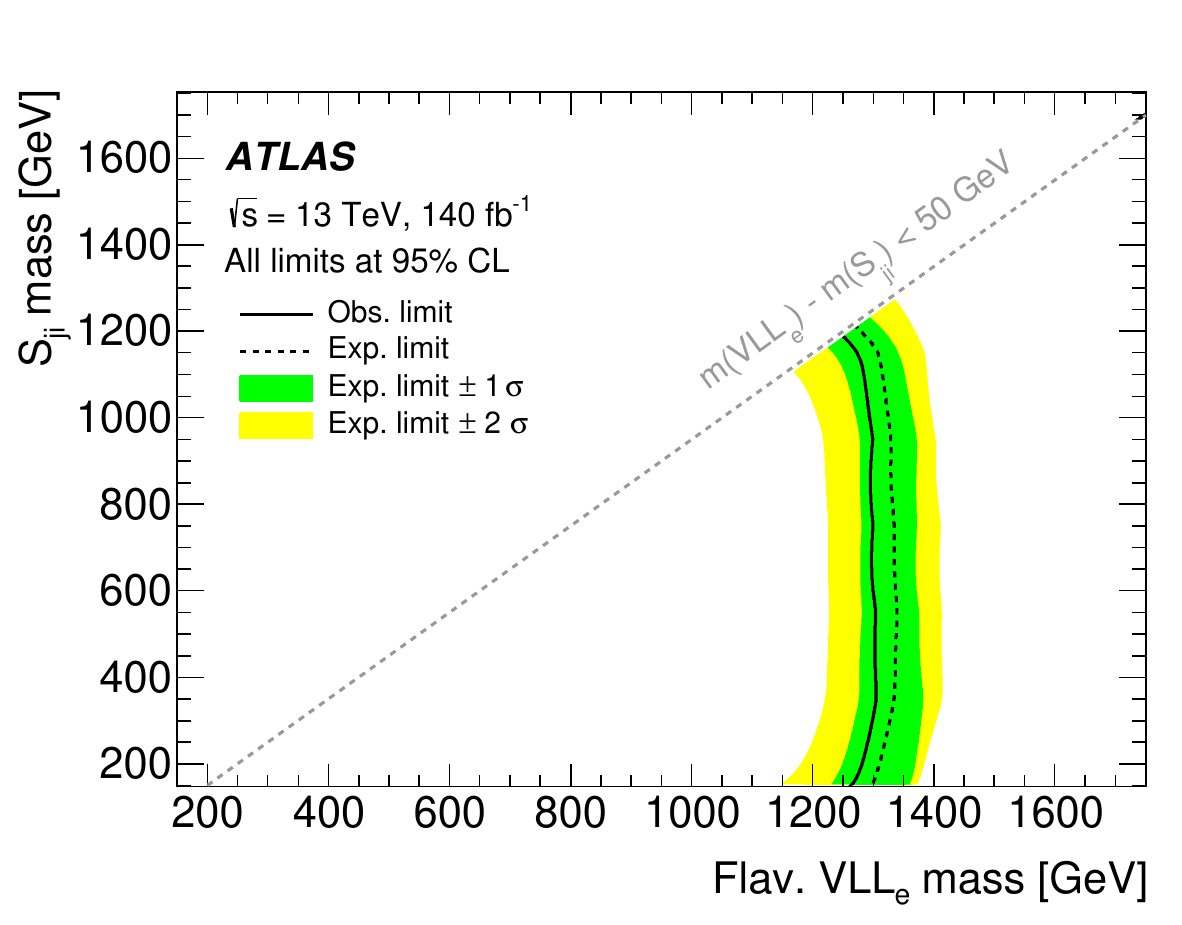}\label{fig:Limits,ModelDep,Fl_VLLe}}
\subfloat[]{\includegraphics[width=0.45\textwidth]{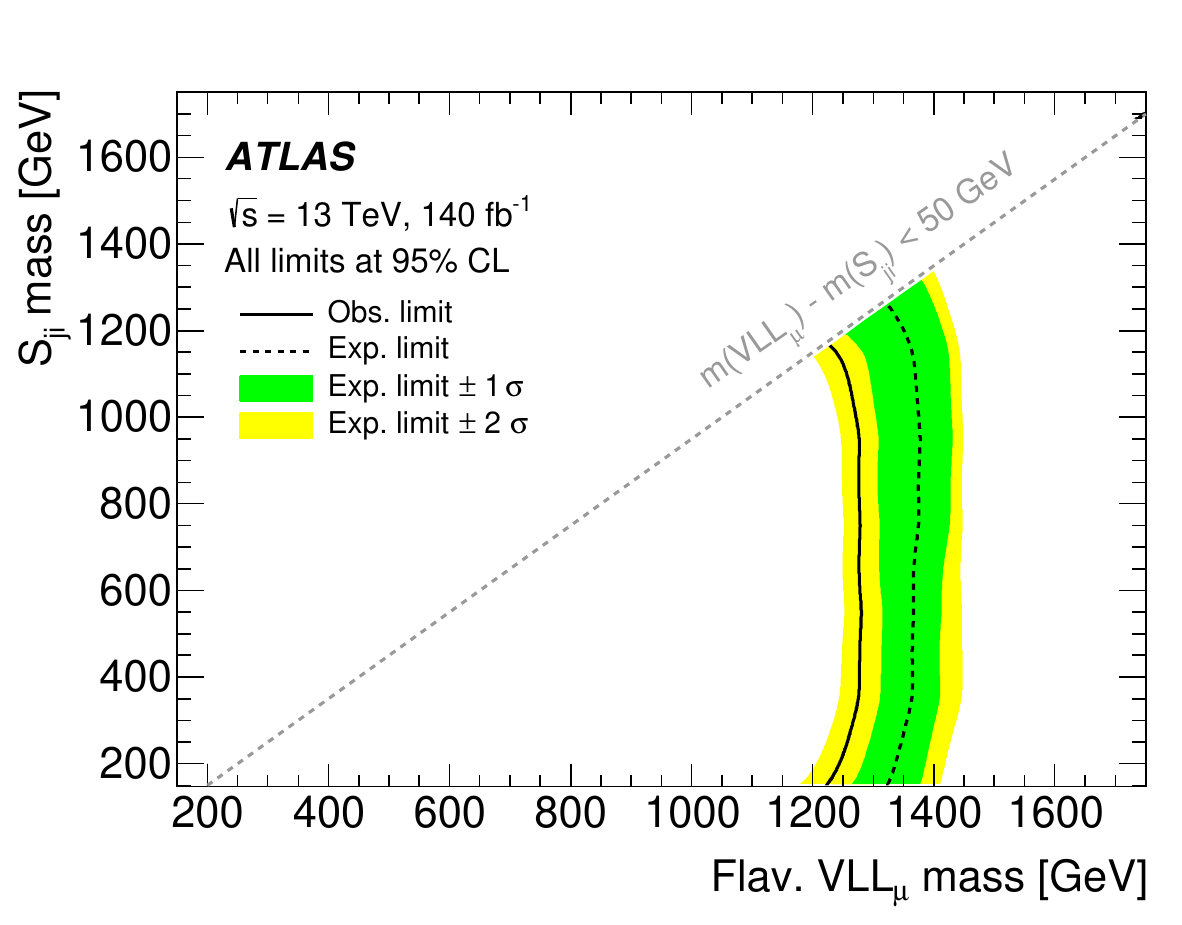}\label{fig:Limits,ModelDep,Fl_VLLmu}}
\caption{
Combined observed and expected limits at 95\% CL on the cross-section ($\sigma$) as a function of the VLL mass for the (a) \vlled, (b) \vllmud, (c) \vlles, (d) \vllmus models, and on the plane of the $S_{ji}$ mass versus the VLL mass for the (e) flavourful \vlle, and (f) flavourful \vllmu models.
The expected limits from the dedicated ATLAS search~\cite{EXOT-2021-31}, considering only the \llll channel, are overlaid in dashed grey lines in (a-d).
The inner and outer bands around the expected limit are the $\pm 1 \sigma$ and $\pm 2 \sigma$ variations including all uncertainties, respectively.
}
\end{figure}

\begin{figure}[!h]
\centering
\subfloat[]{\includegraphics[width=0.45\textwidth]{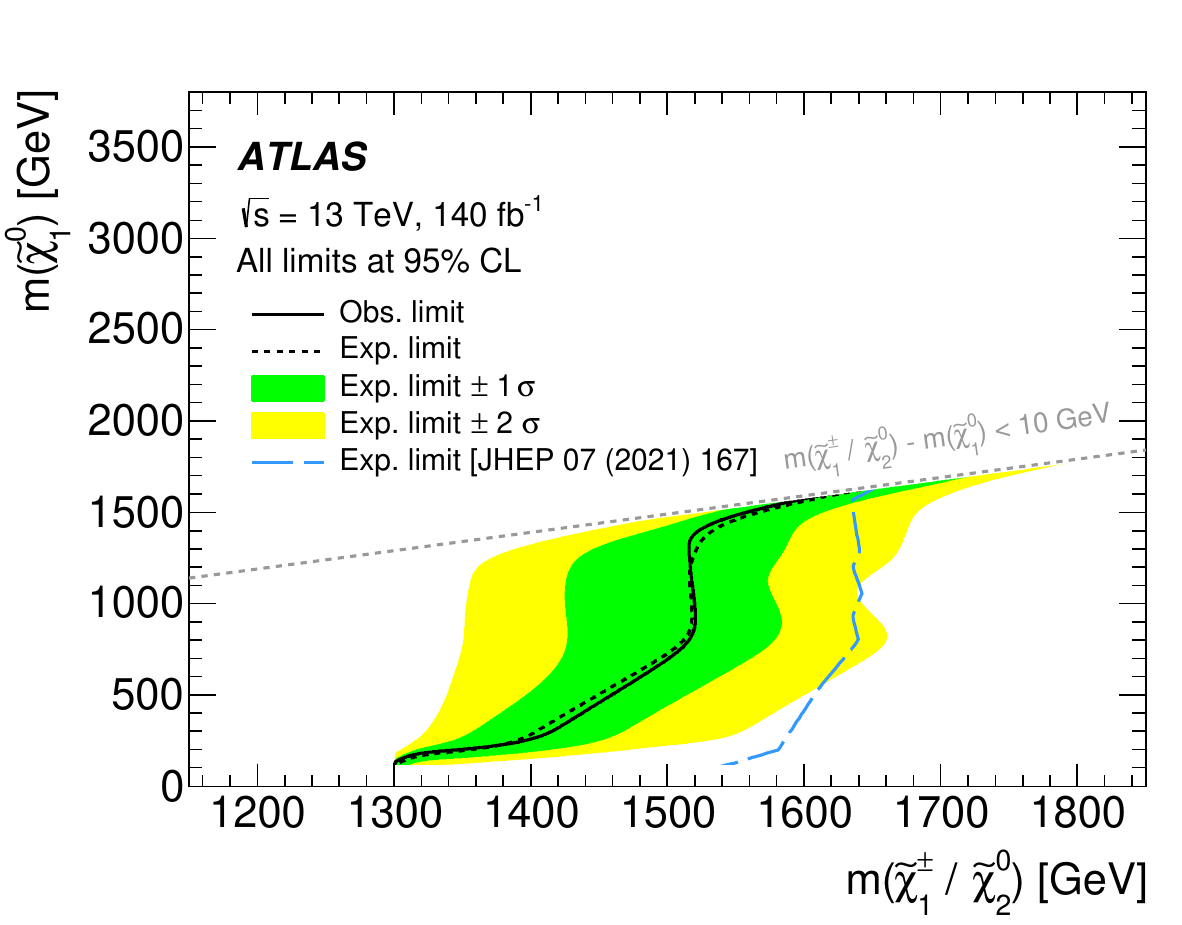}}
\subfloat[]{\includegraphics[width=0.45\textwidth]{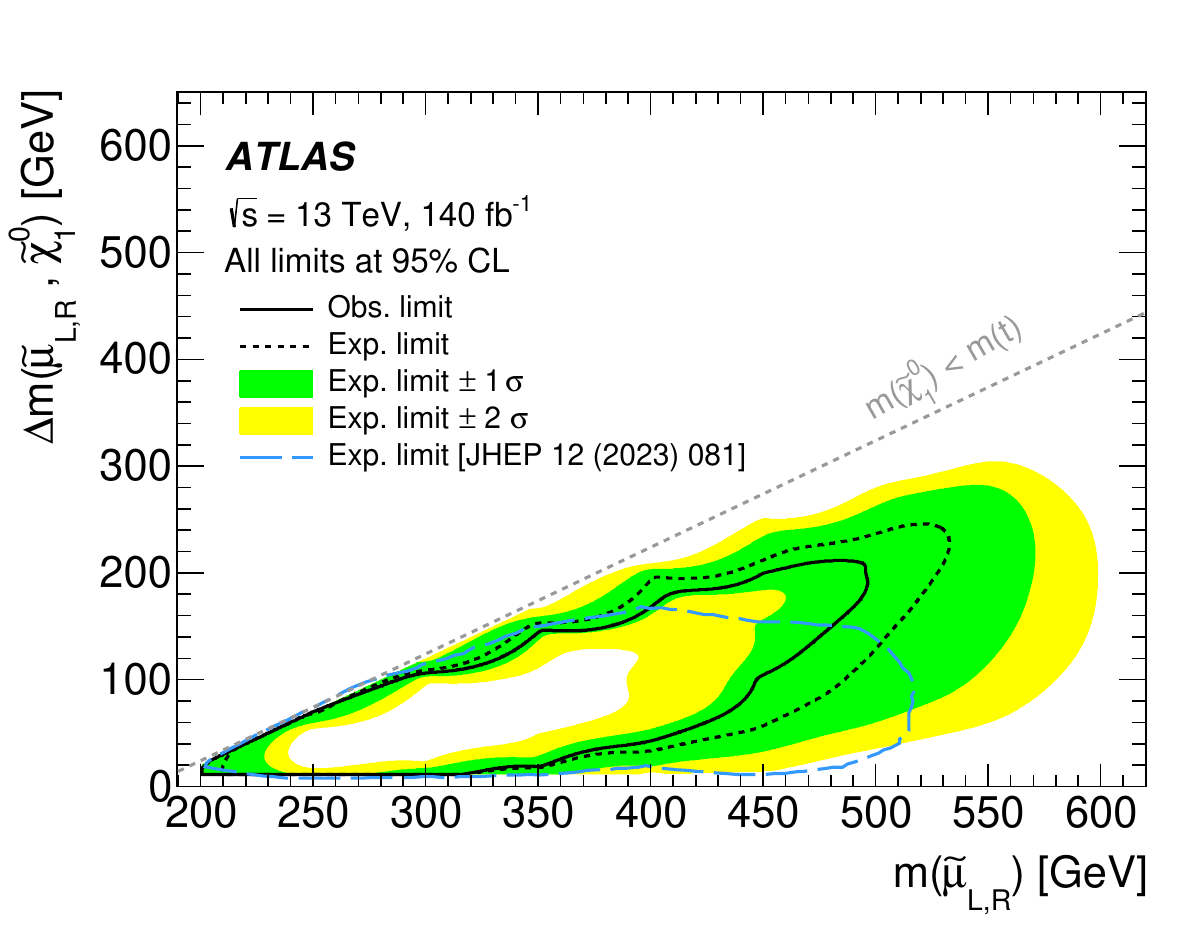}}
\caption{
Combined observed and expected limits at 95\% CL for the (a) wino-like chargino and neutralino, and (b) smuon models, as a function of the masses of $\tilde{\chi}^{\pm}_{1} / \tilde{\chi}^0_2$ and $\tilde{\chi}^0_1$, and $\tilde{\mu}_{\mathrm{L,R}}$ and $\Delta m (\tilde{\mu}_{\mathrm{L,R}}, \tilde{\chi}^0_1)$, respectively.
The expected limits from the dedicated ATLAS searches~\cite{SUSY-2018-02, HDBS-2020-03} are overlaid in dashed grey lines.
The inner and outer bands around the expected limit are the $\pm 1 \sigma$ and $\pm 2 \sigma$ variations including all uncertainties, respectively.
\label{fig:Limits,ModelDep,WinoSmuon}}
\end{figure}


%

%
%
%

\FloatBarrier

\section{Conclusions}
%

A model-independent search for BSM physics with anomaly detection techniques is presented.
The search is based on a data sample of proton--proton collisions recorded at $\sqrt{s}=13~\tev$ by the ATLAS detector during Run 2 of the LHC, corresponding to an integrated luminosity of 140~fb$^{-1}$.
The search is performed in final states with $\geq$ 4 light leptons ($e, \mu$).
The use of anomaly detection optimises the sensitivity of the signal regions to Beyond the Standard Model physics without assuming a specific model. A model-dependent analysis is also performed to demonstrate the sensitivity of the anomaly detection technique with respect to a selection of benchmark models.
The dominant backgrounds originate from $\ZZ$, $\ttdy$, and $\VVV$ production, as well as SM processes with fake leptons, and are estimated from Monte-Carlo simulation and normalised to data in a simultaneous fit.
The data are found to be consistent with the Standard Model predictions and exclusion limits are set on the visible cross section for each model-independent signal region. Additionally, exclusion limits on the mass of vector-like leptons, wino-like charginos and neutralinos, and smuons are set in the model-dependent analysis using the same anomaly detection discriminant as with the model-independent search, with a different event categorisation. The expected exclusion limits are generally competitive to the ones from dedicated searches, in some cases improving the sensitivity in corners of the available phase space. Exclusion limits are set on the flavourful VLL mass for the first time, excluding flavourful \vlle (\vllmu) masses up to 1300 (1280) GeV.
%
%


%

%
%
%
%
%
\section*{Acknowledgements}
%

%

%
%

%
%

We thank CERN for the very successful operation of the LHC and its injectors, as well as the support staff at
CERN and at our institutions worldwide without whom ATLAS could not be operated efficiently.

The crucial computing support from all WLCG partners is acknowledged gratefully, in particular from CERN, the ATLAS Tier-1 facilities at TRIUMF/SFU (Canada), NDGF (Denmark, Norway, Sweden), CC-IN2P3 (France), KIT/GridKA (Germany), INFN-CNAF (Italy), NL-T1 (Netherlands), PIC (Spain), RAL (UK) and BNL (USA), the Tier-2 facilities worldwide and large non-WLCG resource providers. Major contributors of computing resources are listed in Ref.~\cite{ATL-SOFT-PUB-2025-001}.

We gratefully acknowledge the support of ANPCyT, Argentina; YerPhI, Armenia; ARC, Australia; BMWFW and FWF, Austria; ANAS, Azerbaijan; CNPq and FAPESP, Brazil; NSERC, NRC and CFI, Canada; CERN; ANID, Chile; CAS, MOST and NSFC, China; Minciencias, Colombia; MEYS CR, Czech Republic; DNRF and DNSRC, Denmark; IN2P3-CNRS and CEA-DRF/IRFU, France; SRNSFG, Georgia; BMFTR, HGF and MPG, Germany; GSRI, Greece; RGC and Hong Kong SAR, China; ICHEP and Academy of Sciences and Humanities, Israel; INFN, Italy; MEXT and JSPS, Japan; CNRST, Morocco; NWO, Netherlands; RCN, Norway; MNiSW, Poland; FCT, Portugal; MNE/IFA, Romania; MSTDI, Serbia; MSSR, Slovakia; ARIS and MVZI, Slovenia; DSI/NRF, South Africa; MICIU/AEI, Spain; SRC and Wallenberg Foundation, Sweden; SERI, SNSF and Cantons of Bern and Geneva, Switzerland; NSTC, Taipei; TENMAK, T\"urkiye; STFC/UKRI, United Kingdom; DOE and NSF, United States of America.

Individual groups and members have received support from BCKDF, CANARIE, CRC and DRAC, Canada; CERN-CZ, FORTE and PRIMUS, Czech Republic; COST, ERC, ERDF, Horizon 2020, ICSC-NextGenerationEU and Marie Sk{\l}odowska-Curie Actions, European Union; Investissements d'Avenir Labex, Investissements d'Avenir Idex and ANR, France; DFG and AvH Foundation, Germany; Herakleitos, Thales and Aristeia programmes co-financed by EU-ESF and the Greek NSRF, Greece; BSF-NSF and MINERVA, Israel; NCN and NAWA, Poland; La Caixa Banking Foundation, CERCA Programme Generalitat de Catalunya and PROMETEO and GenT Programmes Generalitat Valenciana, Spain; G\"{o}ran Gustafssons Stiftelse, Sweden; The Royal Society and Leverhulme Trust, United Kingdom.

In addition, individual members wish to acknowledge support from CERN: European Organization for Nuclear Research (CERN DOCT); Chile: Agencia Nacional de Investigaci\'on y Desarrollo (FONDECYT 1230812, FONDECYT 1240864, Fondecyt 3240661, Fondecyt Regular 1240721); China: Chinese Ministry of Science and Technology (MOST-2023YFA1605700, MOST-2023YFA1609300), National Natural Science Foundation of China (NSFC - 12175119, NSFC 12275265); Czech Republic: Czech Science Foundation (GACR - 24-11373S), Ministry of Education Youth and Sports (ERC-CZ-LL2327, FORTE CZ.02.01.01/00/22\_008/0004632), PRIMUS Research Programme (PRIMUS/21/SCI/017); EU: H2020 European Research Council (ERC - 101002463); European Union: European Research Council (BARD No. 101116429, ERC - 948254, ERC 101089007), European Regional Development Fund (SMASH COFUND 101081355, SLO ERDF), European Union, Future Artificial Intelligence Research (FAIR-NextGenerationEU PE00000013), Italian Center for High Performance Computing, Big Data and Quantum Computing (ICSC, NextGenerationEU); France: Agence Nationale de la Recherche (ANR-21-CE31-0022, ANR-22-EDIR-0002, ANR-24-CE31-0504-01); Germany: Deutsche Forschungsgemeinschaft (DFG - 469666862, DFG - CR 312/5-2); China: Research Grants Council (GRF); Italy: Istituto Nazionale di Fisica Nucleare (ICSC, NextGenerationEU), Ministero dell'Università e della Ricerca (NextGenEU 153D23001490006 M4C2.1.1, NextGenEU I53D23000820006 M4C2.1.1, NextGenEU I53D23001490006 M4C2.1.1, SOE2024\_0000023); Japan: Japan Society for the Promotion of Science (JSPS KAKENHI JP22H01227, JSPS KAKENHI JP22H04944, JSPS KAKENHI JP22KK0227, JSPS KAKENHI JP24K23939, JSPS KAKENHI JP24KK0251, JSPS KAKENHI JP25H00650, JSPS KAKENHI JP25H01291, JSPS KAKENHI JP25K01023); Norway: Research Council of Norway (RCN-314472); Poland: Ministry of Science and Higher Education (IDUB AGH, POB8, D4 no 9722), Polish National Science Centre (NCN 2021/42/E/ST2/00350, NCN OPUS 2023/51/B/ST2/02507, NCN UMO-2019/34/E/ST2/00393, UMO-2022/47/O/ST2/00148, UMO-2023/49/B/ST2/04085, UMO-2023/51/B/ST2/00920, UMO-2024/53/N/ST2/00869); Portugal: Foundation for Science and Technology (FCT); Spain: Ministry of Science and Innovation (MCIN \& NextGenEU PCI2022-135018-2, MICIN \& FEDER PID2021-125273NB, RYC2019-028510-I, RYC2020-030254-I, RYC2021-031273-I, RYC2022-038164-I); Sweden: Carl Trygger Foundation (Carl Trygger Foundation CTS 22:2312), Swedish Research Council (Swedish Research Council 2023-04654, VR 2021-03651, VR 2022-03845, VR 2022-04683, VR 2023-03403, VR 2024-05451), Knut and Alice Wallenberg Foundation (KAW 2018.0458, KAW 2022.0358, KAW 2023.0366); Switzerland: Swiss National Science Foundation (SNSF - PCEFP2\_194658); United Kingdom: Royal Society (NIF-R1-231091); United States of America: U.S. Department of Energy (ECA DE-AC02-76SF00515), Neubauer Family Foundation.

%
%


%
%
%
%
%
%

%
%
%
\printbibliography
\clearpage
 
\begin{flushleft}
\hypersetup{urlcolor=black}
{\Large The ATLAS Collaboration}

\bigskip

\AtlasOrcid[0000-0002-6665-4934]{G.~Aad}$^\textrm{\scriptsize 104}$,
\AtlasOrcid[0000-0001-7616-1554]{E.~Aakvaag}$^\textrm{\scriptsize 17}$,
\AtlasOrcid[0000-0002-5888-2734]{B.~Abbott}$^\textrm{\scriptsize 123}$,
\AtlasOrcid[0000-0002-0287-5869]{S.~Abdelhameed}$^\textrm{\scriptsize 119a}$,
\AtlasOrcid[0000-0002-1002-1652]{K.~Abeling}$^\textrm{\scriptsize 55}$,
\AtlasOrcid[0000-0001-5763-2760]{N.J.~Abicht}$^\textrm{\scriptsize 49}$,
\AtlasOrcid[0000-0002-8496-9294]{S.H.~Abidi}$^\textrm{\scriptsize 30}$,
\AtlasOrcid[0009-0003-6578-220X]{M.~Aboelela}$^\textrm{\scriptsize 45}$,
\AtlasOrcid[0000-0002-9987-2292]{A.~Aboulhorma}$^\textrm{\scriptsize 36e}$,
\AtlasOrcid[0000-0001-5329-6640]{H.~Abramowicz}$^\textrm{\scriptsize 157}$
% ...
\bigskip
\\

% ...
$^{17}$Department for Physics and Technology, University of Bergen, Bergen; Norway.\\
% ...
$^{30}$Physics Department, Brookhaven National Laboratory, Upton NY; United States of America.\\
% ...
$^{36}$$^{(a)}$Facult\'e des Sciences Ain Chock, Universit\'e Hassan II de Casablanca;$^{(b)}$Facult\'{e} des Sciences, Universit\'{e} Ibn-Tofail, K\'{e}nitra;$^{(c)}$Facult\'e des Sciences Semlalia, Universit\'e Cadi Ayyad, LPHEA-Marrakech;$^{(d)}$LPMR, Facult\'e des Sciences, Universit\'e Mohamed Premier, Oujda;$^{(e)}$Facult\'e des sciences, Universit\'e Mohammed V, Rabat;$^{(f)}$Institute of Applied Physics, Mohammed VI Polytechnic University, Ben Guerir; Morocco.\\
% ...
$^{45}$Physics Department, Southern Methodist University, Dallas TX; United States of America.\\
% ...
$^{49}$Fakult\"{a}t Physik , Technische Universit{\"a}t Dortmund, Dortmund; Germany.\\
% ...
$^{55}$II. Physikalisches Institut, Georg-August-Universit\"{a}t G\"ottingen, G\"ottingen; Germany.\\
% ...
$^{104}$CPPM, Aix-Marseille Universit\'e, CNRS/IN2P3, Marseille; France.\\
% ...
$^{119}$$^{(a)}$New York University Abu Dhabi, Abu Dhabi;$^{(b)}$United Arab Emirates University, Al Ain; United Arab Emirates.\\
% ...
$^{123}$Homer L. Dodge Department of Physics and Astronomy, University of Oklahoma, Norman OK; United States of America.\\
% ...
$^{157}$Raymond and Beverly Sackler School of Physics and Astronomy, Tel Aviv University, Tel Aviv; Israel.\\
% ...
$^{a}$ Also at Affiliated with an institute formerly covered by a cooperation agreement with CERN.\\
$^{b}$ Also at An-Najah National University, Nablus; Palestine.\\
$^{c}$ Also at Borough of Manhattan Community College, City University of New York, New York NY; United States of America.\\
$^{d}$ Also at Center for Interdisciplinary Research and Innovation (CIRI-AUTH), Thessaloniki; Greece.\\
$^{e}$ Also at Centre of Physics of the Universities of Minho and Porto (CF-UM-UP); Portugal.\\
$^{f}$ Also at CERN, Geneva; Switzerland.\\
% ...

\end{flushleft}


%

%
%

%

%
%
%
%

\end{document}